\pdfoutput=1
\documentclass[usegraphicx,useAMS,usenatbib]{mn2e}
\usepackage{tikz,amssymb}
\usepackage{MnSymbol}
\usepackage{fixltx2e}
\usepackage{gensymb}
\newcommand{\kms}{\mbox{km\,s$^{-1}$}}
\def\kmsn{km\,${\rm s}^{-1}$}
\newcommand{\subs}[1]{$_{\rm #1}$}

\def\vsini{$\!${\em v\,}sin{\em i}}
\def\vsinis{$\!${\em v\,}sin{\em i} }
\def\ga{\mathrel{\hbox{\rlap{\hbox{\lower4pt\hbox{$\sim$}}}\hbox{$>$}}}}
\def\la{\mathrel{\hbox{\rlap{\hbox{\lower4pt\hbox{$\sim$}}}\hbox{$<$}}}}
\def\rdd{$\rm{rad\,{d}^{-1}}$}

\title[Magnetic Fields around Moderate Rotators]{Magnetic fields on young, moderately rotating Sun-like stars I: HD~35296 and HD~29615}
\author[I. A. Waite et al.]{I.A.~Waite$^{1}$, S.C.~Marsden$^{1}$, B.D.~Carter$^{1}$, P. Petit$^{2,3}$, J.-F.~Donati$^{2,3}$, S.V.~Jeffers$^{4}$, \and S. Boro Saikia$^{4}$ \thanks
{E-mail:waite@usq.edu.au} \\
${^1}$Computational Engineering and Science Research Centre, University of Southern Queensland, Toowoomba, 4350, Australia    \\
${^2}$Universit\'{e}de Toulouse, UPS-OMP, Institut de Recherche en Astrophysique et Plan\'{e}tologie, F-31400 Toulouse, France 	\\
${^3}$CNRS, Institut de Recherche en Astrophysique et Plan\'{e}tologie, 14 Avenue Edouard, Belin, F-31400, Toulouse, France \\
$^{4}$Institut f\"{u}r Astrophysik, Georg-August-Universit\"{a}t G\"{o}ttingen, Friedrich-Hund-Platz 1, 37077 G\"{o}ttingen, Germany
}

\begin{document}

\date{}

\pagerange{\pageref{firstpage}--\pageref{lastpage}} \pubyear{2002}

\maketitle

\label{firstpage}

\begin{abstract}

Observations of the magnetic fields of young solar-type stars provide a way to investigate the signatures of their magnetic activity and dynamos. Spectropolarimetry enables the study of these stellar magnetic fields and was thus employed at the T\'{e}lescope Bernard Lyot and the Anglo-Australian Telescope to investigate two moderately rotating young Sun-like stars, namely HD~35296 (V119~Tau, HIP~25278) and HD~29615 (HIP~21632). The results indicate that both stars display rotational variation in chromospheric indices consistent with their spot activity, with variations indicating a probable long-term cyclic period for HD~35296. Additionally, both stars have complex, and evolving, large-scale surface magnetic fields with a significant toroidal component. High levels of surface differential rotation were measured for both stars. For the F8V star HD 35296 a rotational shear of $\Delta\Omega$ = 0.22$^{+0.04}_{-0.02}$ \rdd\ was derived from the observed magnetic profiles. For the G3V star HD 29615 the magnetic features indicate a rotational shear of $\Delta\Omega$ = 0.48$_{-0.12}^{+0.11}$ \rdd, while the spot features, with a distinctive polar spot, provide a much lower value of $\Delta\Omega$ of 0.07$_{-0.03}^{+0.10}$ \rdd. Such a significant discrepancy in shear values between spot and magnetic features for HD 29615 is an extreme example of the variation observed for other lower-mass stars. From the extensive and persistent azimuthal field observed for both targets it is concluded that a distributed dynamo operates in these moderately rotating Sun-like stars, in marked contrast to the Sun's interface-layer dynamo.

\end{abstract}


\section[Introduction]{Introduction}
\label{introduction}

Observations of the distribution of magnetic fields offer insights into how magnetic dynamos are governed by physical properties such as mass, rotation and age, with younger Sun-like stars providing proxies for understanding the history of the Sun's activity and dynamo. Today we can readily observe the starspots and surface magnetic fields of young Sun-like stars, to reconstruct their surface magnetic topologies and differential rotation. Multi-epoch observations of stellar magnetic regions also offers the prospect of revealing how magnetic activity cycles can emerge in stars like the Sun. Such observations can allow us to study the relative role of two possible dynamo theories, namely the widely accepted Sun's ``shell'' dynamo that operates at the interface layer between the radiation and convection zone and the ``distributed'' dynamo thought to be operating in young, active stars where the dynamo is present throughout the convection zone \citep[e.g.][]{Brandenburg89,Moss95,Donati03a,Brown10}. While the large-scale toroidal magnetic field is understood to be buried deep inside the Sun, it has been observed on a range of rapidly rotating solar-type stars through the presence of strong unipolar surface azimuthal magnetic fields \citep[e.g.][]{Donati03a,Petit04a,Marsden06}. \citet{Petit08} concluded that the rotation period threshold where the surface toroidal field begins to dominate over the poloidal field is $\approx$12~d. Thus, rotation plays an important role in the generation of the magnetic field and the complexity of these fields.

The study of these complex magnetic fields on Sun-like stars have been greatly advanced using the technique of Zeeman Doppler imaging (ZDI) \citep{Semel89,Donati90,DonatiBrown97,Donati03a}. ZDI has been performed for a small sample of single, rapidly rotating early G-type stars such as HD~171488 (V889 Her; SpType: G2V) \citep{Marsden06,Jeffers08,Jeffers11} and HD~141943 (SpType: G2) \citep{Marsden11a, Marsden11b}. Differential rotation, given its likely importance to the operation of the dynamo, is a key measurement. These stars have shown significant levels of differential rotation using both brightness and magnetic features. \citet{Jeffers08} observed HD~171488 and measured $\Delta\Omega$ = 0.52 $\pm$ 0.04 \rdd\ using spot features while $\Delta\Omega$ = 0.47 $\pm$ 0.04 \rdd\ was measured using magnetic features. \citet{Jeffers11} in their more recent study measured slightly lower values of $\Delta\Omega$ $\sim$ 0.4 \rdd (Stokes $\it{I}$) and $\sim$ 0.415-0.45 \rdd (Stokes $\it{V}$). These levels of differential rotation are similar to an earlier study by \citet{Marsden06} of $\Delta\Omega$ = 0.402 $\pm$ 0.044 \rdd\ (using spot features). \citet{Marsden11b} observed HD~141943 and measured $\Delta\Omega$ = 0.24 $\pm$ 0.03 \rdd\ using spot features with $\Delta\Omega$ = 0.45 $\pm$ 0.08 \rdd\ using magnetic features on their 2010 dataset (and slightly lower with $\Delta\Omega$ = 0.36 $\pm$ 0.09 \rdd\ for their 2007 data). Conversely, slowly rotating stars such as HD~190771 (SpType: G2V), with a \vsinis of 4.3 $\pm$ 0.5 \kms, also exhibit measurable differential rotation with $\Delta\Omega$ = 0.12 $\pm$ 0.03 \rdd\ using magnetic features \citep{Petit08}. In addition to relatively high levels of differential rotation, all three stars exhibited complex magnetic topologies with large-scale poloidal and toroidal fields, each with significant higher order components beyond simple dipole fields.

\citet{Barnes05} first reported the link between differential rotation and effective temperature, and consequently convective turnover time. They were able to develop a correlation between rotational shear and effective temperature for a small sample of stars and determined that as a star's effective temperature increases so does the rotational shear; a trend consistent with the recent theoretical calculations by \citet{Kitchatinov11}. Additionally, a steep rise in differential rotation for late F-/early G-type stars appears related to the rapid shallowing of the convection zone \citep[e.g.][]{Marsden11b}. \citet*{Kuker11} demonstrate that the extreme levels of rotational shear observed for stars such as HD~171488 can only be explained with a shallow convection zone. Thus, the depth of the convection zone must have a major role in the rotational profile of a solar-type star.  

In this study, two moderately rotating, young Sun-like stars were selected based on their similar rotational velocities and sizes, enabling comparisons of their magnetic fields and differential rotations. HD~35296 (SpType: F8V) is a $1.06_{-0.05}^{+0.06}$ $M_\odot$ \citep*{Holmberg09} star with an estimated photospheric temperature 6170K \citep{Casagrande11} while HD~29615 (SpType: G3V) is a 0.95 $M_\odot$, 1.0 $R_\odot$ star \citep{AllendePrieto99}. HD~35296 has a \vsinis of 15.9 \kms \citep{Ammler12} while HD~29615 has a \vsinis of $\sim$18 \kms \citep{Waite11b}. The aim is to map the surfaces of these two stars with a view to measuring the rotational shear and magnetic field structures. This will enable further comparisons with stars of similar mass but different rotational velocities. This is paper I of two papers focusing on the complex nature of the magnetic field topologies of young moderately rotating Sun-like stars as proxies for the evolution of the young Sun. Paper II will focus on the infant Sun EK~Draconis (Waite et al. {\it in preparation}). This paper is part of the BCool\protect\footnote{http://bcool.ast.obs-mip.fr} collaboration investigating the magnetic activity of low-mass stars \citep[e.g.][]{Marsden14}.

\section[Fundamental Parameters of HD~35296 and HD~29615]{Fundamental Parameters of the stars}

The fundamental parameters for the two targets, HD~35296 and HD~29615, are shown in Table \ref{parameters}. Some of these parameters such as projected rotational velocity (\vsini), radial velocity and inclination were measured as part of the imaging process, as explained in Sect. \ref{ImageReco}, while other parameters have been taken from the literature.

\subsection{HD~35296}
HD~35296 (V1119 Tau) is a F8V star \citep{Montes01}. It has a parallax of 69.51 $\pm$ 0.38 milliarcseconds (mas) \citep{Van_Leeuwen07} giving a distance of 14.39 $\pm$ 0.08 parsecs (pc). \citet{Samus09} identified HD~35296 as a BY Draconis-type variable star. The range of age estimates of this star in the literature is quite large, with \citet{Holmberg09} estimating an age of 3.3~Gyr, with some authors listing this star as young as 20 Myr or as old as 7.5~Gyr \citep[e.g.][]{Chen01,Barry88}. The equivalent width of the Li~{\sc i} 670.78~nm line is 99~m\AA, correcting for the nearby 670.744~nm Fe~{\sc i} line using the same factor developed by \citet{Soderblom93a,Soderblom93b}. This is consistent with that measured by \citet{Takeda05} of 94.3 m\AA. \citet{Li98} argued that HD~35296 is a member of the Taurus-Auriga star forming region that may have reached the Zero-Age Main Sequence (ZAMS). When placing this star on the theoretical isochrones of \citet*{Siess00}, the age of this star is between 20 Myr to 50 Myr. Hence it is likely that this star is quite youthful.


\subsection{HD~29615}

HD~29615 (HIP~21632) is a G3V star \citep{Torres06}. The $\it{HIPPARCOS}$ space mission measured a parallax of 18.27 $\pm$ 1.02 mas \citep{Van_Leeuwen07}, giving a distance of $54.7_{-2.9}^{+3.2}$ pc. \citet{Zuckerman04} proposed that HD~29615 was a member of the Tucana/Horologium Association indicating an age of $\sim$30 Myr. \citet{Waite11b} detected a magnetic field on this star, along with a varying emission equivalent width for the H$\alpha$ line in the range from $\sim$370 m\AA\ to $\sim$500 m\AA\ demonstrating the presence of a very active, and variable, chromosphere. 


\begin{table*}
\begin{center}
\caption{The parameters used to produce the maximum-entropy image reconstructions of the two targets, including surface differential rotation measurements. Except otherwise indicated, parameters have been determined by this study.} 
\label{parameters}
\begin{tabular}{llll}
\hline
Parameter         					& HD~35296	& HD~29615	                \\
\hline
Spectral Type					& F8V$^{1}$		  & G3V$^{2}$				\\
Equatorial Period			   	& 3.48 $\pm$ 0.01~d~$^{3}$   & 2.34$_{-0.05}^{+0.02}$ d~$^{4}$		\\
Inclination Angle 			 	& 65 $\pm$ 5$\degree$~$^{5}$        & 65$_{-10}^{+5}$$\degree$       		\\
Projected Rotational Velocity, \vsini 	 	& 15.9 $\pm$ 0.1 \kmsn    & 19.5 $\pm$ 0.1 \kmsn  	 	\\
Photospheric Temperature, T\subs{phot} 	 	& 6170~K $^{6}$         & 5820 $\pm$ 50~K $^{7}$		\\
Spot Temperature, T\subs{spot} 			& --         		& 3920 K $^{8}$			\\
Radial Velocity, v\subs{rad}	 	  	& 38.1 $\pm$ 0.1 \kmsn    & 19.33 $\pm$ 0.1 \kmsn   		\\
Stellar radius 					& 1.10 R$_{\odot}$$^{9}$ & 1.0 $R_\odot$$^{10}$ 		\\ 
Stellar mass					& $1.06_{-0.05}^{+0.06}$ $M_\odot$$^{9}$ & 0.95 $M_\odot$$^{10}$  \\
Age 					 	& 30-50 Myr $^{12}$	& 30 Myr${^{11}}$		\\
Convection zone depth$^{12}$ $R_{\star}$($R_{\odot}$)	& 0.178 (0.201) & 0.252 (0.252)		\\
Stokes $\it{I}$: $\Omega$\subs{eq} \rdd		&  --  			& 2.68$_{-0.02}^{+0.06}$  	\\
Stokes $\it{I}$: $\Delta\Omega$	\rdd		&  --  			& 0.07$_{-0.03}^{+0.10}$ 	\\
Stokes $\it{V}$: $\Omega$\subs{eq} \rdd		&  1.804 $\pm$ 0.005   	& 2.74$_{-0.04}^{+0.02}$  	 	\\
Stokes $\it{V}$: $\Delta\Omega$	\rdd		&  0.22$^{+0.04}_{-0.02}$ 	& 0.48$_{-0.12}^{+0.11}$  	\\
Epoch of zero phase (MHJD)$^{13}$		& 54133.871035 (2007) & 55165.011060 (2009)	       \\
						& 54496.094272 (2008) &				\\
\hline
\end{tabular}
\end{center}
$^{1}$ \citet{Montes01},
$^{2}$ \citet{Torres06},
$^{3}$ Using Stokes $\it{V}$ data, 
$^{4}$ Using Stokes $\it{I}$ data, 
$^{5}$ The inclination angle of HD~35296 was based on the most recent value of P~sin$\theta$ = 3.9 calculated by \citet{Ammler12}.
$^{6}$ \citet{Casagrande11},
$^{7}$ Using the bolometric corrections of \citet{Bessell98},
$^{8}$ Using the relationship between photospheric and spot temperature provided by \citet{Berdyugina05},
$^{9}$ based on the values found in \citet*{Holmberg09},
$^{10}$ \citet{AllendePrieto99},
$^{11}$ \citet{Zuckerman04}, 
$^{12}$ Determined from the stellar evolution models of \citet{Siess00}, 
$^{13}$ Modified Heliocentric Julian Date (MHJD) = HJD - 24000000.0.\\
\end{table*}

\section[Observations and Analysis]{Observations and Analysis}
HD~35296 was observed at the T\'{e}lescope Bernard Lyot (TBL - Observatoire Pic du Midi, France) in January/February, 2007 and again in January/February, 2008 using the high-resolution spectropolarimeter NARVAL. HD~29615 was observed at the Anglo-Australian Telescope (AAT - New South Wales, Australia) in November/December, 2009 using the high-resolution spectropolarimeter SEMPOL. Journals of observations are given in Tables \ref{spectroscopy_log_TBL} and \ref{spectroscopy_log_AAT}.

\subsection{High Resolution Spectropolarimetric Observations from the TBL}
\label{Narval}
NARVAL is a bench mounted, cross-dispersed \'{e}chelle spectrograph, fibre-fed from a Cassegrain-mounted polarimeter unit and is similar in function to the Semel Polarimeter (SEMPOL) used at the AAT, which is described in Sect. \ref{sempol_exp}. NARVAL has a mean pixel resolution of 1.8 \kms\ per pixel with a spectral coverage from $\sim$ 370~nm to 1048~nm with a resolution of $\sim$65000 spanning 40 grating orders (orders \#22 to \#61). NARVAL consists of one fixed quarter-wave retarder sandwiched between two rotating half-wave retarders and coupled to a Wollaston beamsplitter \citep{Auriere03}. Observations in circular polarization (Stokes {\it V}) consist of a sequence of four exposures. After each of the exposures, the half-wave Fresnel Rhomb of the polarimeter is rotated so as to remove any systematic effects due to: variations in the optical throughput, CCD inhomogeneities, terrestrial and stellar rotation, temporal variability and instrumental polarization signals from the telescope and the polarimeter \citep[e.g.][]{Semel93,Carter96,Donati03a}. Initial reduction was completed using the dedicated pipeline reduction software {\small LibreES{\textsc p}RIT} (\'{E}chelle Spectra Reduction: an Interactive Tool), which is based on the algorithm developed by \citet{Donati97}. \citet{Silvester12} proved that NARVAL is a very stable instrument, with resolution and signal-to-noise ratio (SNR) being constant over the four years of data studies from 2006-2010.

\subsection{High Resolution Spectropolarimetric Observations from the AAT}
\label{sempol_exp}

High resolution spectropolarimetric data were obtained from the AAT using the University College of London \'{E}chelle Spectrograph (UCLES) and SEMPOL \citep{Semel93,Donati97,Donati03a}. The detector used was the deep depletion EEV2 CCD with 2048 x 4096 13.5~$\mu$m square pixels. UCLES was used with a 31.6 gr/mm grating covering 46 orders (orders \#~84 to \#~129). The central wavelength was 522.002~nm with full wavelength coverage from 438~nm to 681~nm. The dispersion of $\sim$0.004958~nm at order  \#~129 gave a resolution of approximately 71000. The mean resolution for the AAT spectra were determined to be 1.689~\kms\ pixel$^{-1}$. The operation of SEMPOL is similar to that explained in Sect. \ref{Narval}. After each exposure, the half-wave Fresnel Rhomb is rotated between +45$\degree$ and -45$\degree$ so as to remove any systematic effects. Initial reduction was completed using ES{\sc p}RIT developed for SEMPOL \citep{Donati97}. SEMPOL is not as efficient as NARVAL, hence the SNR is lower, even though the AAT is a 3.9-m telescope whereas the TBL is a 2-m telescope. In addition, the spectral range of SEMPOL is smaller therefore there are fewer spectral lines available to extract the magnetic signature from the star's light.



\begin{table}
\begin{center}
\caption{Journal of spectropolarimetric observations of HD~35296 using the TBL.} 
\label{spectroscopy_log_TBL}
\begin{tabular}{cccc}
\hline
UT Date &  UT           & Exp. Time${^1}$ & Mean SNR for   	  \\ 
        &  middle       &    (sec)        & Stokes $\it{V}$    	  \\
	& 		& 		  & LSD profiles \\
\hline
2007 Jan 24 & 21:08:45 & 1$\times$300 & ---$^{2}$ \\
2007 Jan 26 & 21:10:23 & 4$\times$300 & 19762  \\
2007 Jan 27 & 20:18:17 & 4$\times$600 & 39439  \\
2007 Jan 29 & 19:29:15 & 4$\times$600 & 38358  \\
2007 Feb 02 & 21:17:29 & 4$\times$300 &  4223  \\
2007 Feb 03 & 20:15:21 & 4$\times$300 & 29940  \\
2007 Feb 04 & 20:06:35 & 4$\times$300 & 25137  \\
2007 Feb 08 & 20:30:51 & 4$\times$300 & 25639  \\
\hline
2008 Jan 19 & 20:01:46 & 4$\times$300 & 17208  \\
2008 Jan 20 & 21:45: 1 & 4$\times$300 & 20581  \\
2008 Jan 21 & 19:42:27 & 4$\times$300 & 24563  \\
2008 Jan 21 & 20:53:24 & 4$\times$200 & 22146  \\
2008 Jan 22 & 20:09:11 & 4$\times$300 & 21913  \\
2008 Jan 23 & 20:10:31 & 4$\times$300 & 21825  \\
2008 Jan 24 & 20:39:32 & 4$\times$300 & 13296  \\
2008 Jan 25 & 20:15:17 & 4$\times$300 & 19723  \\
2008 Jan 26 & 20:14:43 & 4$\times$300 & 21148  \\
2008 Jan 27 & 20:30:50 & 4$\times$300 & 21245  \\
2008 Jan 28 & 21:56:43 & 4$\times$300 & 21245  \\
2008 Jan 29 & 20:50:19 & 4$\times$300 & 25946  \\
2008 Feb 02 & 20:45:15 & 4$\times$300 & 24375  \\
2008 Feb 04 & 20:20:19 & 4$\times$300 & 27669  \\
2008 Feb 05 & 20:21:18 & 4$\times$300 & 22909  \\
2008 Feb 06 & 21:11: 7 & 4$\times$300 & 25579  \\
2008 Feb 09 & 20:34:17 & 4$\times$300 & 24555  \\
2008 Feb 10 & 19:13:58 & 4$\times$300 & 29354  \\
2008 Feb 11 & 20:39:11 & 4$\times$300 & 25414  \\
2008 Feb 12 & 20:45:32 & 4$\times$300 & 27269  \\
2008 Feb 13 & 20:39:30 & 4$\times$300 & 24182  \\
2008 Feb 14 & 21:00: 8 & 4$\times$300 & 24386  \\
2008 Feb 15 & 20:37: 6 & 4$\times$300 & 28120  \\
\hline 
\end{tabular}

\end{center}
{${^1}$ Each sequence consists of four sub-exposures with each sub-exposure being 300 s (for example)},  \\
{${^2}$ Only one Stokes $\it{I}$ spectrum was taken on this night. The SNR, as measured for order number 41, was 272}.  \\
\end{table}


\begin{table}
\begin{center}
\caption{Journal of spectropolarimetric observations of HD~29615 using the AAT.} 
\label{spectroscopy_log_AAT}
\begin{tabular}{cccc}
\hline
UT Date &  UT  		&  Exp. Time${^1}$ 	& Mean SNR for     	\\
        &  middle       &    (sec)              & Stokes $\it{V}$  	\\
	& 		& 		  	& LSD profiles \\
\hline
2009 Nov 25 & 11:03:28 &  4$\times$750 & 5567 \\
2009 Nov 25 & 13:04:01 &  4$\times$750 & 6337 \\
2009 Nov 25 & 16:29:33 &  4$\times$750 & 4908 \\
2009 Nov 27 & 10:23:36 &  4$\times$750 & 6171 \\
2009 Nov 27 & 13:28:05 &  4$\times$750 & 6978 \\
2009 Nov 27 & 16:28:02 &  4$\times$750 & 5610 \\
2009 Nov 28 & 10:10:47 &  4$\times$750 & 3821 \\
2009 Nov 28 & 13:13:08 &  4$\times$750 & 5971 \\
2009 Nov 28 & 16:15:29 &  4$\times$750 & 4280 \\
2009 Nov 29 & 10:08:37 &  400 + 3$\times$750 & 5785 \\
2009 Nov 29 & 13:23:08 &  4$\times$750 & 4604 \\
2009 Nov 29 & 16:23:23 &  4$\times$750 & 3493 \\
2009 Nov 30 & 11:05:48 &  4$\times$750 & 3283 \\
2009 Nov 30 & 16:18:04 &  4$\times$750 & 3713 \\
2009 Dec 01 & 11:02:55 &  4$\times$750 & 6353 \\
2009 Dec 01 & 14:13:39 &  4$\times$750 & 4961 \\
2009 Dec 02 & 10:36:24 &  4$\times$750 & 6158 \\
2009 Dec 02 & 13:04:01 &  4$\times$750 & 6176 \\
2009 Dec 02 & 16:03:39 &  4$\times$750 & 4651 \\
2009 Dec 03 & 10:27:35 &  4$\times$750 & 4228 \\
2009 Dec 03 & 13:33:32 &  4$\times$750 & 6312 \\
\hline
\end{tabular}
\end{center}
{${^1}$ Each sequence consists of four sub-exposures with each sub-exposure being 750 s (for example)}. \\
\end{table}


\subsection{Spectropolarimetric Analysis}
\label{SpAnalysis}
Preliminary processing involved subtracting the bias level and using a nightly master flat combining typically 20 flat field exposures. Each stellar spectrum was extracted and wavelength calibrated against a Thorium-Argon lamp.  

After using {\small LibreES{\textsc p}RIT} (TBL) or {\small  ES{\textsc p}RIT} (AAT), the technique of Least Squares Deconvolution (LSD) was applied to the reduced spectra. LSD combines the information from many spectral lines to produce a single line profile, thereby providing an enormous multiplex gain in the SNR \citep*[e.g.][]{DonatiCameron97,Donati97,Kochukhov10}. A G2 line mask created from the Kurucz atomic database and ATLAS9 atmospheric models \citep{Kurucz93} was used to compute the average line profile for both stars. Tables \ref{spectroscopy_log_TBL} and \ref{spectroscopy_log_AAT} lists the SNR for each individual Stokes $\it{V}$ LSD profile.

In order to correct for the minor instrumental shifts in wavelength space due to atmospheric temperature or pressure fluctuations, each spectrum was shifted to match the centroid of the Stokes $\it{I}$ LSD profile of the telluric lines contained in the spectra, as was performed by \citet{Donati03a} and \citet{Marsden06}. Further information on LSD can be found in \citet{Donati97} and \citet{Kochukhov10}.

\section{Chromospheric activity indicators}
\label{chromo_activity}

Chromospheric activity can be determined using the Ca $\textsc{ii}$ H \& K, Ca $\textsc{ii}$ Infrared Triplet (IRT) and H$\alpha$ spectral lines. The two Ca $\textsc{ii}$ H \& K absorption lines are the most widely used optical indicators of chromospheric activity. There have been a number of long-term monitoring studies of the variation of Ca $\textsc{ii}$ H \& K in solar-type stars spanning several decades. One study is the Mount Wilson Ca H \& K project \citep[e.g.][]{Wilson78,Duncan91}. From this survey, differential rotation and solar-like activity cycles have been inferred on a number of solar-type stars \citep[e.g.][]{Baliunas85,Donahue92,Donahue93,Donahue94}. 

The source functions of the Ca $\textsc{ii}$ H \& K lines are collisionally controlled hence are very sensitive to electron density and temperature. The  Ca $\textsc{ii}$ IRT lines share the upper levels of the H \& K transitions and are formed in the lower chromosphere \citep[e.g.][]{Montes04}. The H$\alpha$ spectral line is also collisionally filled in as a result of the higher temperatures and is formed in the middle of the chromosphere and is often associated with plages and prominences \citep[e.g.][]{Thatcher93,Montes04}.  

The Ca $\textsc{ii}$ H \& K and Ca $\textsc {ii}$ Infrared Triplet (IRT) spectral lines were observed for HD~35296 but not for HD~29615 as the spectral range of SEMPOL does not extend far enough into the respective regions to permit monitoring of these diagnostic lines. 

\subsection{TBL Activity Indices for HD~35296}
\label{TBL_Activity_Indices}
\citet{Morgenthaler12} previously reported that the spectra in the region of the Ca $\textsc{ii}$ H \& K are not well normalised by the pipe-line reduction due to the dense distribution of photospheric lines. However, we did not adopt their method as a number of tests showed that simply removing the overlapping section of the order provided commensurate results as renormalising the spectra.

The N$_{Ca \textsc{ii}HK}$-index for HD~35296 was determined using the method as explained in \citet{Wright04}. The resulting N$_{Ca~\textsc {ii} HK}$-index was converted to match the Mount Wilson S-values \citep{Duncan91} using the transformation shown in Eqn. \ref{MountWilson}. 

\begin{equation}
	\label{MountWilson}	
	\ S\textrm{-}index = \frac{C_{1} H + C_{2} K}{C_{3} V_{HK} + C_{4} R_{HK}}+C_{5}  
\end{equation}

where H and K is the flux determined in the line cores from the two triangular bandpasses with a full-width at half maximum (FWHM) of 0.1~nm. Two 2~nm-wide rectangular bandpasses R$_{HK}$ and V$_{HK}$, centred on 400.107 and 390.107~nm respectively, were used for the continuum flux in the red and blue sides of the H and K lines. The transformation coefficients were determined by \citet{Marsden14} by matching 94 stars that were common to both the TBL database and that of the Mount Wilson project. These coefficients were: C$_{1}$ = 12.873, C$_{2}$ = 2.502, C$_{3}$ = 8.877, C$_{4}$ = 4.271 and C$_{5}$ = 1.183 $\times$ 10$^{-3}$.

Two further activity indices were determined using the H$\alpha$ spectral line and the Ca \textsc {ii} IRT lines. The continuum was checked against that of the continuum of the synthetic normalised spectrum from the POLLUX database \citep{Palacios10}. In both cases,the continuum matched reasonably well hence was not corrected. The TBL H$\alpha$-index, based on \citet{Gizis02}, was determined using Eqn. \ref{Halpha}. 

\begin{equation}
	\label{Halpha}
		\ N_{H\alpha}\textrm{-}index = \frac{F_{H\alpha}}{V_{H\alpha}+R_{H\alpha}}
\end{equation}
where F$_{H\alpha}$ is the flux determined in the line cores from the triangular bandpass with a FWHM of 0.2~nm. Two 0.22~nm-wide rectangular bandpasses V$_{H\alpha}$ and R$_{H\alpha}$, centred on 655.885 and 656.730~nm respectively, were used for the continuum flux in the red and blue sides of the H$\alpha$ line. The TBL CaIRT-index, following \citet{Petit13}, was determined using using Eqn. \ref{CaIRT} .

\begin{equation}
	\label{CaIRT}
		\ N_{CaIRT}\textrm{-}index = \frac{\sum F_{IRT}}{V_{IRT} + R_{IRT}}
\end{equation}
where $\sum F_{IRT}$ is the total flux measured in the line cores of the three spectral lines, 849.8023, 854.2091 and 866.2141~nm using triangular bandpasses with a FWHM of 0.1~nm. Two 0.5~nm-wide rectangular bandpasses V$_{IRT}$ and R$_{IRT}$, centred on 847.58 and 870.49~nm respectively, were used for the continuum flux in the red and blue sides of the Ca \textsc {ii} IRT spectral lines.

\begin{figure*}
\begin{center}
\includegraphics[trim = 0cm 1.5cm 0cm 0cm, scale=0.87, angle=0]{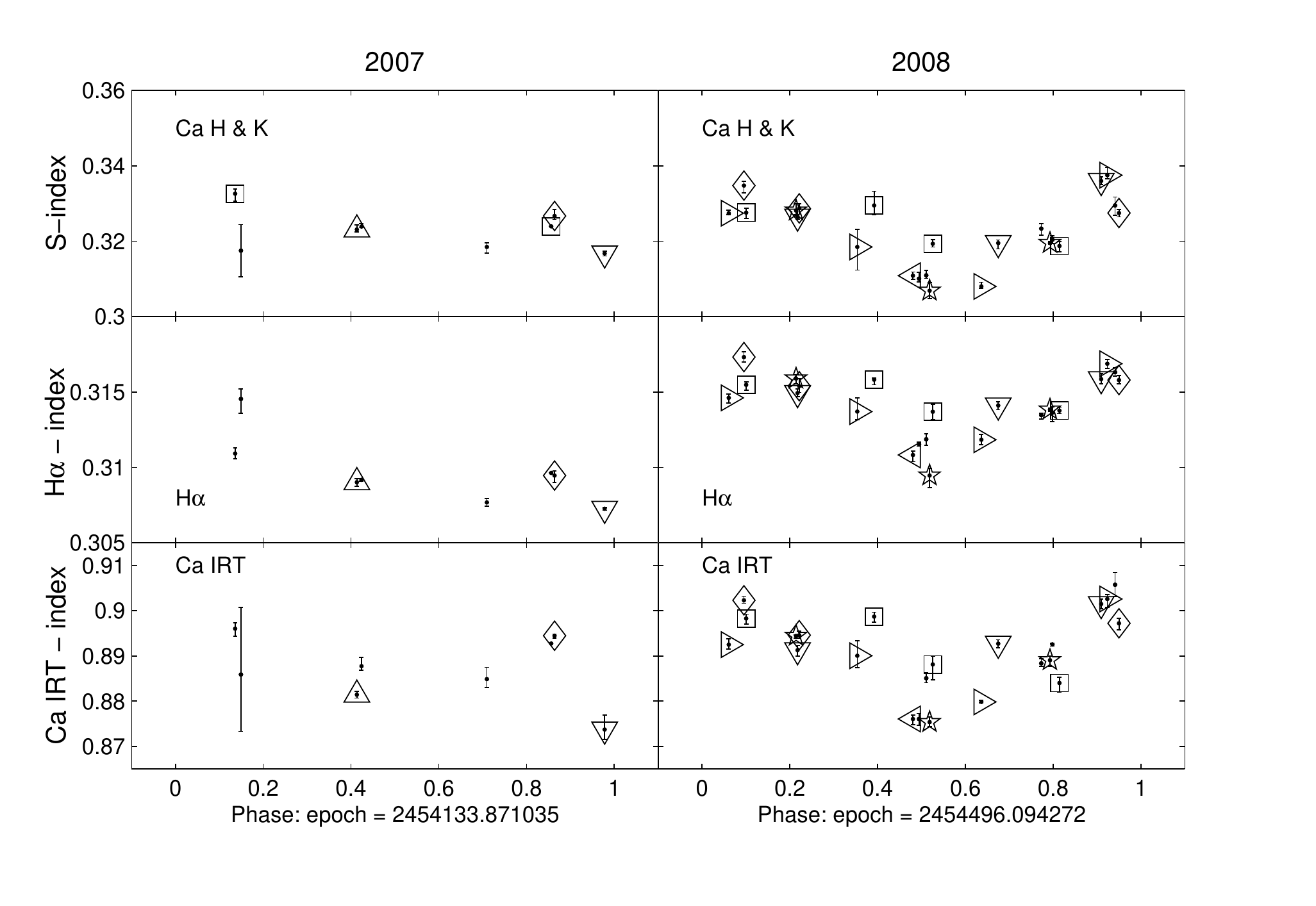}
\caption{The variation in the Ca $\textsc{ii}$ H \& K lines (upper), H$\alpha$ (middle) and Ca $\textsc{ii}$ IRT (lower) spectral lines during 2007 (left) and 2008 (right) for HD~35296. The index was measured for each spectrum, then combined in sets of four, coinciding with a cycle of 4 sub-exposures as explained in Sect. \ref{Narval}. The average of these four indices was determined with the error bar being the minimum and maximum values for that set. The rotational period used was 3.48~d. The observing run covered $\sim$4.3 rotations in 2007 and $\sim$7.7 rotations in 2008. The symbols represent the respective rotations: . only = first, $\diamond$ = second, $\square$ = third, $\triangle$ = fourth, $\triangledown$ = fifth, $\triangleleft$ = sixth, $\triangleright$ = seventh and $\medstar$ = eighth rotation.}

\label{S-index} 
\end{center}
\end{figure*}

Fig. \ref{S-index} shows the variation in the Ca $\textsc{ii}$ H \& K lines, Ca $\textsc{ii}$ IRT and H$\alpha$ spectral lines during 2007 (left panels) and 2008 (right panels) for HD~35296. The epoch for the 2007 data was set to the middle of the observation run whereas the epoch of the 2008 data was set to 104 stellar rotations later on and was chosen to be close to the middle of the 2008 observation run, as explained in Sect. \ref{MagneticImages}. The index was measured for each spectrum, then combined in sets of four, coinciding with a cycle of 4 sub-exposures as explained in Sect. \ref{Narval}. The average of these four indices was determined with the error bar being the minimum and maximum values for that set. The data from 2007 showed limited variation, either due to the reduced coverage or a more homogeneous chromosphere. In contrast, in 2008, HD~35296 exhibited greater modulation of the S-index, CaIRT-index and H$\alpha$-index while the average remained very similar with an S-index for 2007 of 0.323 $\pm$ 0.008 and for 2008 of 0.322 $\pm$ 0.016. This is consistent, within error limits, with the average S-index of 0.308 $\pm$ 0.07 from the Mount Wilson survey \citep{Duncan91}. Using the transformation values from \citet{Rutten84}, {\it Log R$^\prime$$_{HK}$} was determined to be -4.45 $\pm$ 0.03. This was an average value of the S-index over the observing runs with the error based on the minimum and maximum values. Table \ref{activity_indices} show the average S-index, N$_{H\alpha}$ and N$_{CaIRT}$ indices for 2007 and 2008.  Each value is the average over the observing run with the error showing the variation during this time, most likely as a result of modulation due to the rotation of the star.

\begin{table}
\begin{center}
\caption{HD~35296: Average activity indices using the Ca \textsc{ii} H\&K, H$\alpha$ and Ca \textsc{ii} IRT spectral lines. The error estimate is based on the range of values during each observing run.} 
\label{activity_indices}
\begin{tabular}{lcc}
\hline
Index         		& 2007			& 2008	                \\
\hline
S-index$^{1}$		& 0.323 $\pm$ 0.008	& 0.322 $\pm$ 0.016  \\
N$_{H\alpha}$		& 0.310 $\pm$ 0.004	& 0.314 $\pm$ 0.004  \\
N$_{CaIRT}$		& 0.887 $\pm$ 0.011	& 0.891 $\pm$ 0.015  \\
\hline
\end{tabular}
\end{center}
${^1}$The resulting N$_{Ca \textsc{ii}HK}$-index was converted to match the Mount Wilson S-values \citep{Duncan91} using the transformation shown in  Eqn. \ref{MountWilson}. \\
\end{table}

\subsection{AAT \texorpdfstring{H$\alpha$}{} activity index for HD~29615}
\label{AAT_Halpha}

\begin{figure}
\begin{center}
\includegraphics[trim = 0cm 0.0cm 5.2cm 0cm, clip=true, scale=0.60, angle=0]{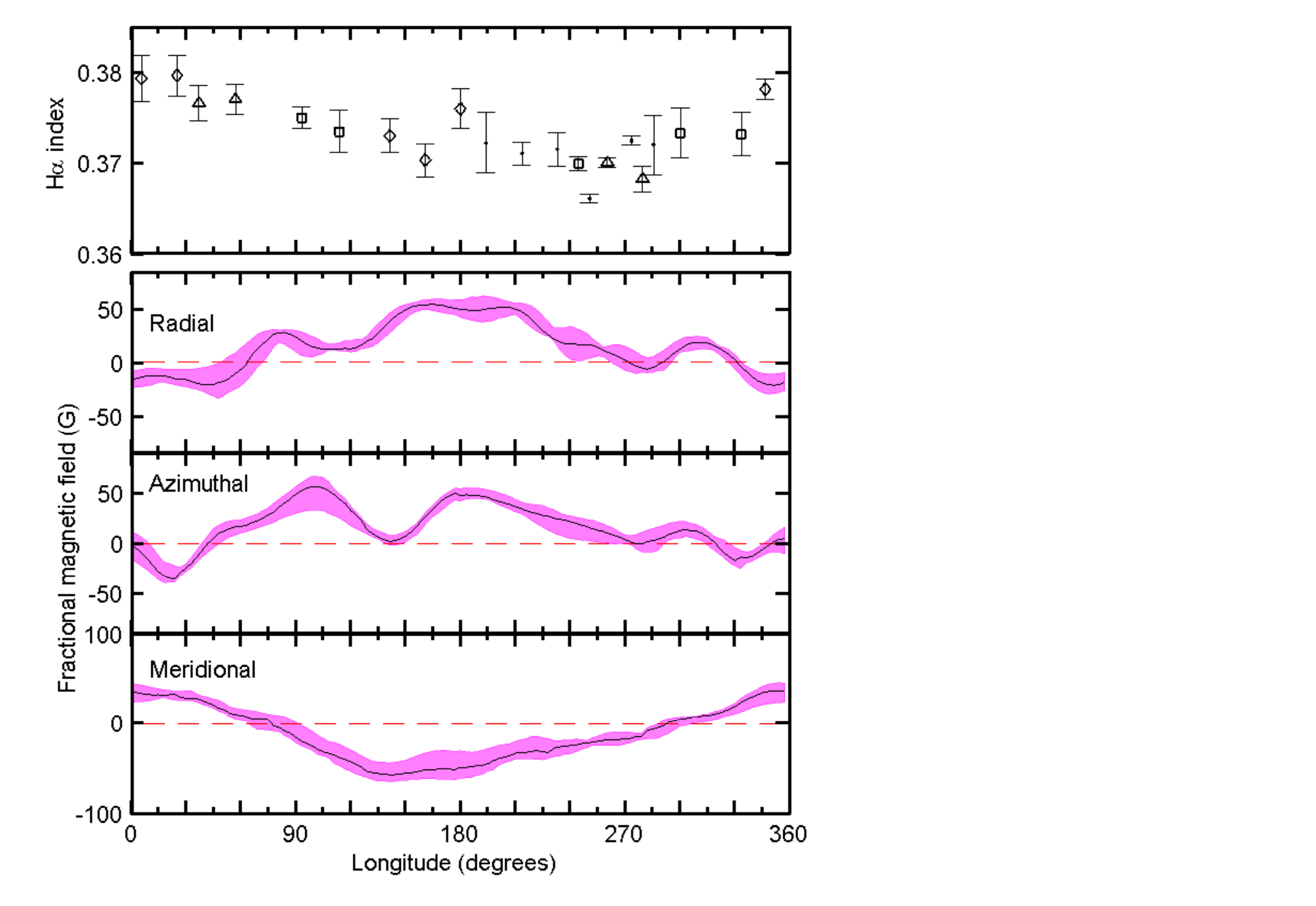}
\caption{The top panel represents the variation in the H$\alpha$ index as a function of longitude for HD~29615 (see Sect. \ref{AAT_Halpha}). The observing run covered $\sim$3.5 rotations. The symbols represent the respective rotations: . = first, $\diamond$ = second, $\square$ = third and $\triangle$ = fourth rotation. The bottom three panels represent the variation in magnetic field strength for each configuration as a function of longitude (see Sect. \ref{dis_chromo_act}). The shaded regions show the variation in the strength at each longitude as determined by varying \vsini, inclination angle, $\Omega_{eq}$ and $\Delta\Omega$ within the limits of their respective accuracy. The Epoch was set to MHJD = 55165.011060 with a rotational period of 2.34~d.} 
\label{hip21632_halpha}
\end{center}
\end{figure}

The H$\alpha$ spectral line was used as a chromospheric indicator for HD~29615. A similar approach, as explained in \ref{TBL_Activity_Indices} was applied to the AAT data. The overlapping sections of the AAT spectra were removed and the continuum checked against a synthetic spectrum obtained from the POLLUX database, In all cases, the continuum near the H$\alpha$ spectral line closely matched the synthetic spectra. Eqn. \ref{Halpha} was then used to determine the H$\alpha$ index for HD~29615. Fig. \ref{hip21632_halpha} shows how this H$\alpha$-index varied over the rotation period of HD~29615. The error bars indicate the range of the measurement during a cycle of 4 sub-exposures, as explained in Sect. \ref{Narval}. This supports the variable nature of the mid-level chromosphere, first observed by \citet{Waite11b}.

\section{Image Reconstruction}
\label{ImageReco}

Surface images of these moderately rotating stars were generated through the inversion of a time series of LSD profiles of Stokes $\it{I}$ (brightness images) or Stokes $\it{V}$ (magnetic images). The Zeeman Doppler imaging code used was that of \citet{Brown91} and \citet{DonatiBrown97}. Because this inversion process is an ill-posed problem where an infinite number of solutions is possible when fitting to the noise level, this code implements the \citet{Skilling84} maximum-entropy optimization that produces an image with the minimum amount of information required to fit the data to the noise level. 

\subsection{Brightness image of HD~29615}

Brightness images were not reconstructed for HD~35296 as the \vsini\ of 15.9 \kms\ is below the usual limit for Doppler imaging. In contrast, the observed intensity profiles for HD~29615 contained sufficient information for both mapping and a measurement of differential rotation using spot features.

A two-temperature model, one being the temperature of the quiet photosphere while the second is that of the cool spots, has been used to reconstruct the brightness image for HD~29615. Synthetic Gaussian profiles were used to represent the profiles of both the spot and photosphere. \citet{Unruh95} showed that the use of these synthetic profiles provides brightness maps that are very similar to those maps created by using profiles from slowly rotating stars of commensurate temperature. Hence, many authors now adopt this approach \citep[e.g.][]{Petit04b,Marsden05,Marsden06,Marsden11a,Marsden11b,Waite11a}. Using the relationship between photospheric and spot temperature provided by \citet{Berdyugina05}, the spot temperature of HD~29615 was estimated to be 3920~K. 

The imaging code was used to establish the values of a number of basic parameters, including the star's projected rotational velocity, \vsini, and radial velocity, v$_{rad}$. This was achieved by systematically varying each parameter in order to minimize the reduced-$\chi^2$ ($\chi^2_r$) value \citep[e.g.][]{Marsden05,Jeffers08}. These key stellar parameters were determined with initially v$_{rad}$ and then \vsini. This sequence was repeated each time additional parameters, such as inclination or differential rotation parameters, were modified as a result of the imaging process. 

\begin{figure*}
\begin{center}
\includegraphics[scale=0.65, angle=-90]{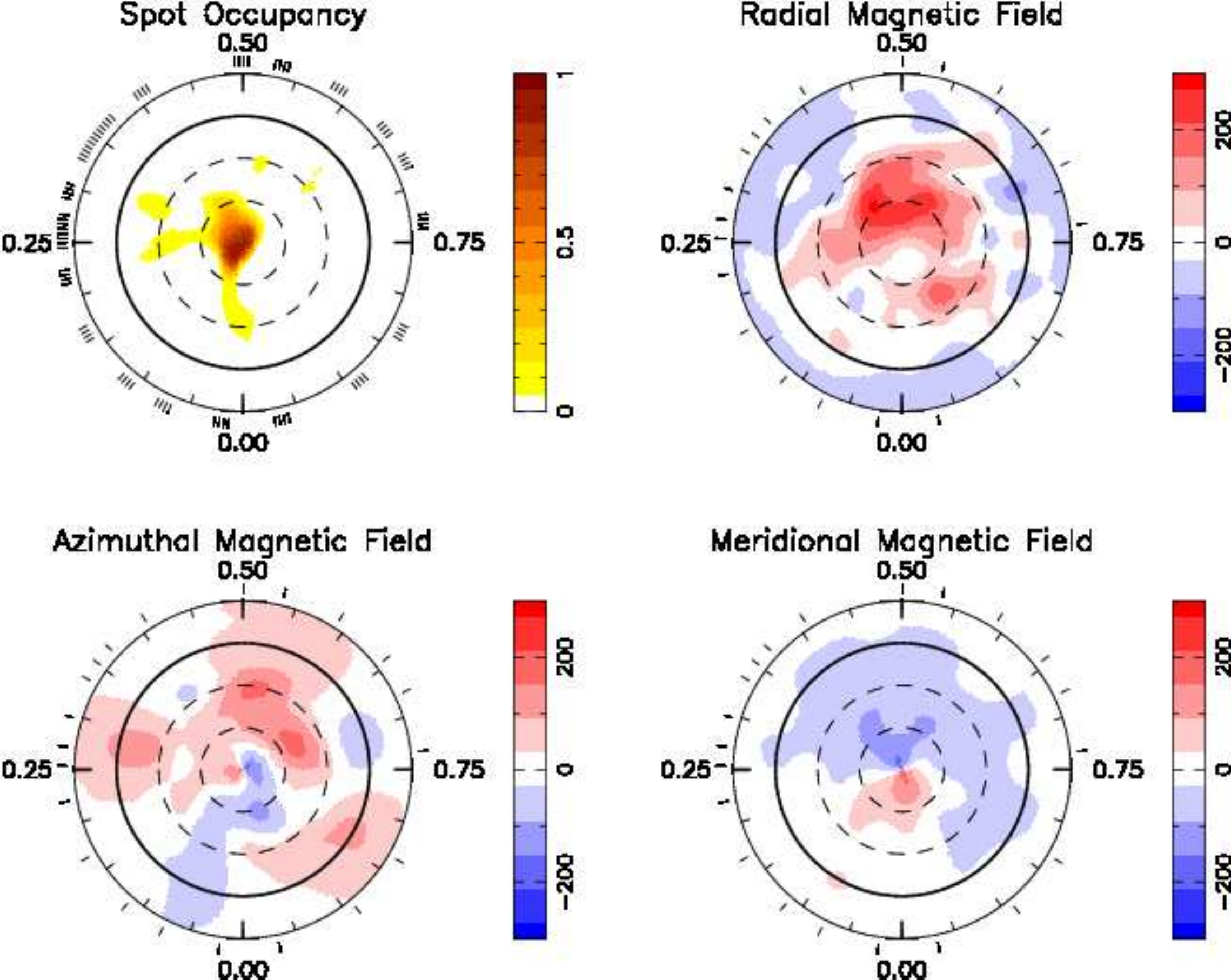}
\caption{The maximum-entropy brightness and magnetic image reconstructions for HD~29615. These maps are polar projections extending down to -30${^o}$. The bold lines denote the equator and the dashed lines are +30${^o}$ and +60${^o}$ latitude parallels. The radial ticks indicate the phases at which this star was observed and the scale of the magnetic images is in Gauss. Differential rotation, as measured using the Stokes $\it{I}$ and $\it{V}$ profiles respectively, has been incorporated into the reconstruction of the maps. The spot map has a spot coverage of 2.6 per cent while the global magnetic field strength is 81.6~G. The Epoch was set to MHJD = 55165.011060 with a rotational period of 2.34~d.}
\label{hip21632_maps} 
\end{center}
\end{figure*}

\begin{figure*}
\begin{center}
\includegraphics[scale=0.9, angle=0]{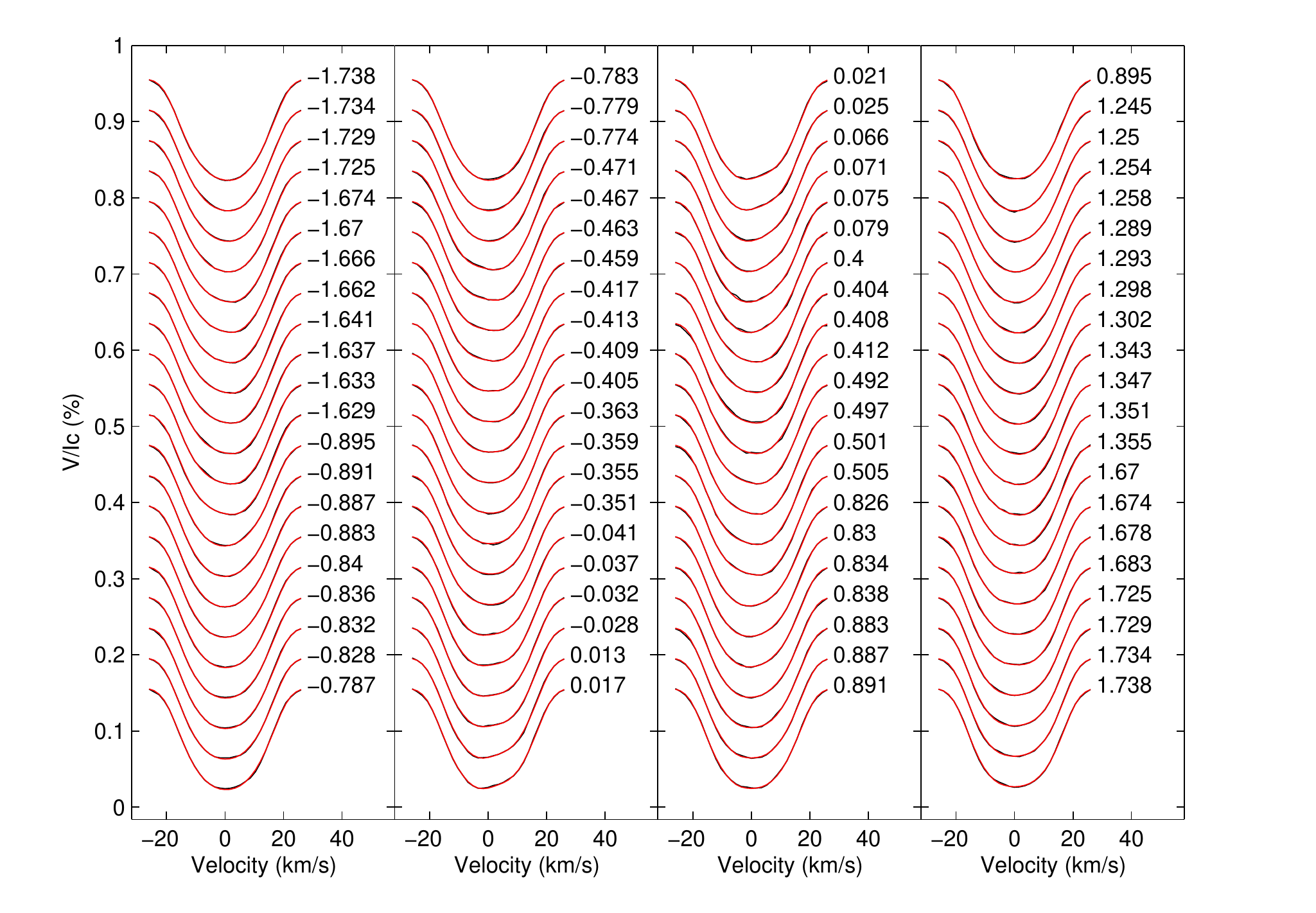}
\caption{The maximum-entropy fits to the Stokes $\it{I}$ LSD profiles for HD~29615 with the measured surface differential rotation incorporated into the analysis. The red lines represent the modelled lines produced by the Doppler imaging process whereas the black lines represent the actual observed LSD profiles. Each successive profile has been shifted down by 0.020 for graphical purposes. The rotational cycle at which the observation took place are indicated to the right of each profile. The Epoch was set to MHJD = 55165.011060 with a rotational period of 2.34~d.}
\label{hip21632_I_fits} 
\end{center}
\end{figure*}

\begin{figure}
\begin{center}
\includegraphics[scale=0.40, angle=0]{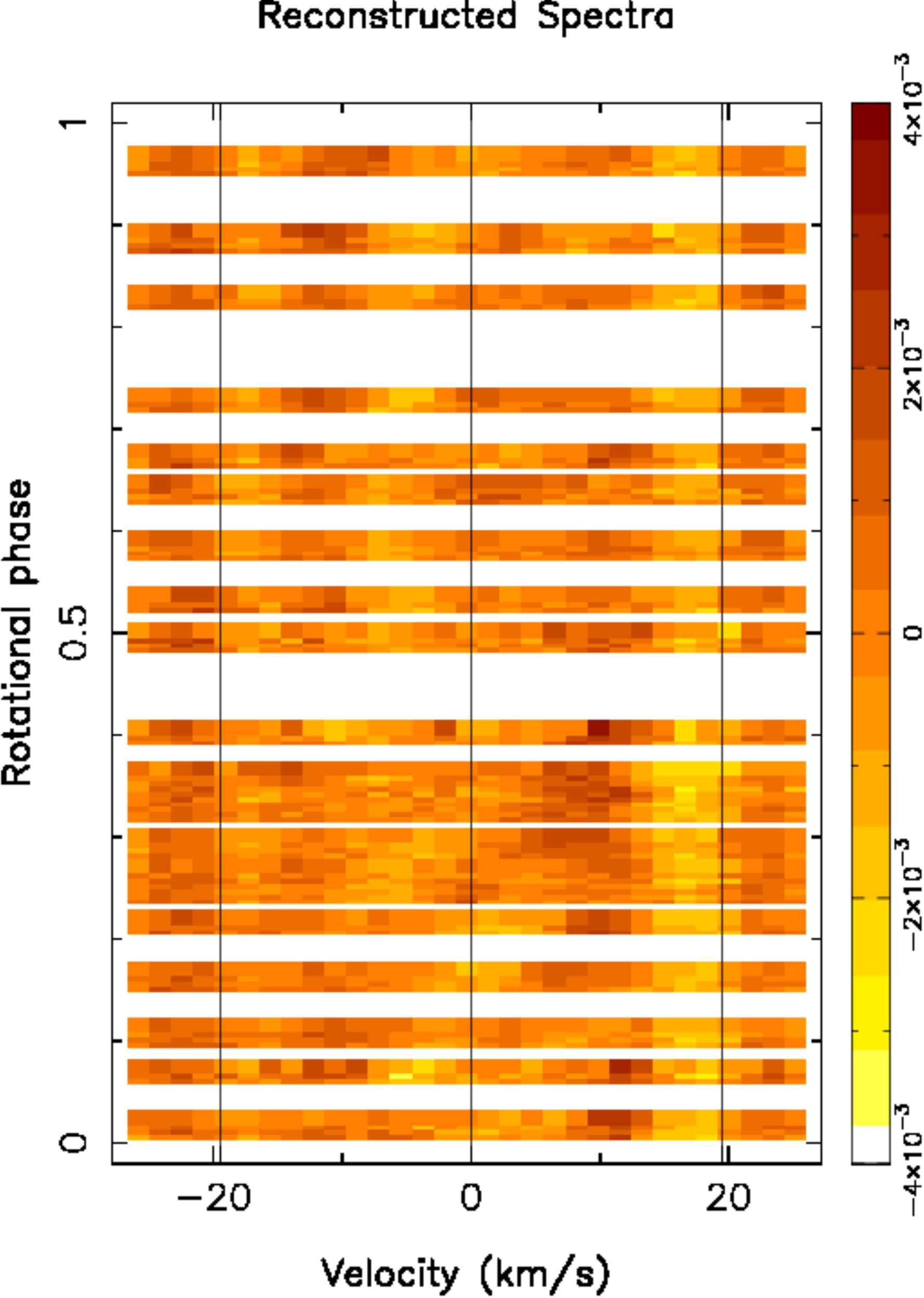}
\caption{Dynamic spectrum of the residual plot for the Stokes $\it{I}$ LSD profiles for HD~29615. Each LSD profile was normalised and subtracted from the associated normalised model profile.}
\label{hip21632_residuals}
\end{center}
\end{figure}

The inclination angle for HD~29615 was estimated using the bolometric corrections of \citet*{Bessell98} and was confirmed using the imaging code. This value was determined to be 65$^{+5}_{10}$$\degree$. The minimum angle of $\sim$ 50$\degree$ assumed a stellar radius of 1.2 R$_\odot$ \citep{Messina10}. A linear limb-darkening coefficient of 0.62 was used \citep{Sing10}. The full set of parameters that gave the minimum $\chi^2_r$ value of 0.45 are shown in Table \ref{parameters} and were adopted when producing the final Doppler imaging map as shown in the top left panel of Fig. \ref{hip21632_maps}. A $\chi^2_r$ value less than one is possible as the SNR calculated for the LSD Stokes $\it{I}$ profiles are underestimated \citep[e.g.][]{Petit04b,Marsden11a}. Nevertheless, this has no impact on the final brightness maps. The maximum-entropy fits to the Stokes $\it{I}$ LSD profiles for HD~29615 with the measured surface differential rotation incorporated into the analysis is shown in Fig. \ref{hip21632_I_fits}.  Deviations of the model profile from the observed profile are shown in the dynamic spectrum displayed in Fig. \ref{hip21632_residuals}. To produce this dynamic spectrum, each observed LSD profile was normalised, then subtracted from the associated normalised model profile with the residual profile being plotted as the dynamic spectrum. The scale of this dynamic spectrum is $\pm$~4~$\times$~10$^{-3}$ or $\approx$ 0.4 per cent of the intensity of the line profile. The small residuals indicate that the majority of the large-scale features can be accurately modelled. However, there is a consistent mis-match between the modelled and original data as evidenced by a brighter band near the red wing of the dynamic spectrum. This indicates that the modelled data has not incorporated information from the wings of the original line profiles. This is advantageous in that modelling of that part of the line profile may have led to spurious banding in the resulting image \citep{Unruh95}.  


\begin{figure*}
\begin{center}
\includegraphics[trim = 0.5cm 0.0cm 4.0cm 0cm, clip=true, scale=1.30, angle=0]{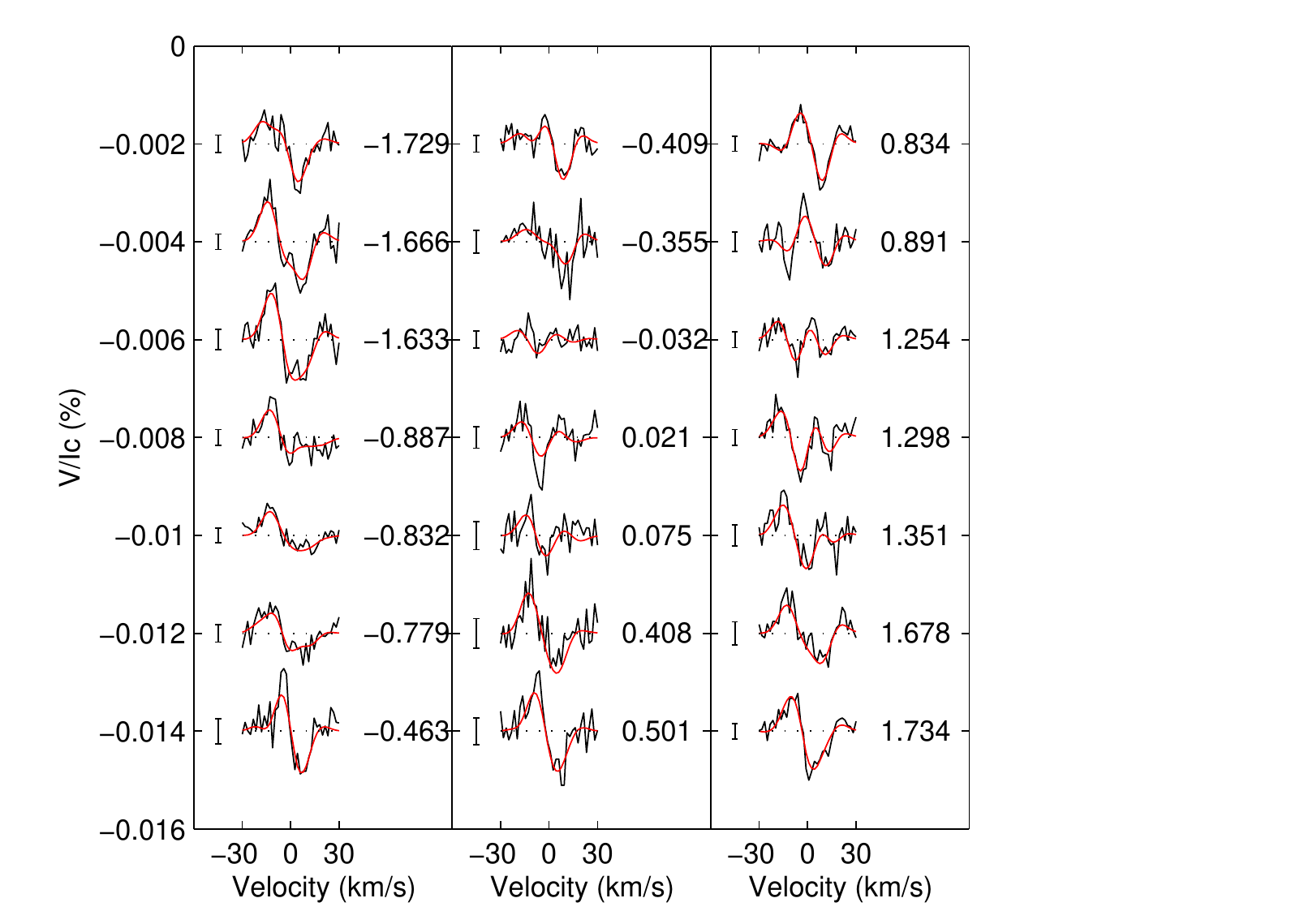}
\caption{The maximum-entropy fits to the Stokes $\it{V}$ LSD profiles for HD~29615 with the measured surface differential rotation incorporated into the analysis. The black lines represent the observed Zeeman signatures, while the red line represent the modelled lines. Each successive profile has been shifted down by 0.002 for graphical purposes. The minimum $\chi^2_r$ value was set to 1.25. The rotational phases at which the observations took place are indicated to the right of each profile. The error bars to the left of each profile are $\pm$ 0.5 $\sigma$. The Epoch was set to MHJD = 55165.011060 with a rotational period of 2.34~d.}
\label{hip21632_V_fits} 
\end{center}
\end{figure*}

\subsection{Magnetic images}
\label{MagneticImages}
The magnetic topology was determined for both HD~35296 and HD~29615 using ZDI. ZDI requires high SNR data as the polarization signature is typically less than 0.1 per cent of the total light intensity \citep{Donati97}. The Stokes $\it{V}$ data was used to reconstruct radial, azimuthal and meridional fields coinciding with the three vector fields in spherical geometry. The modelling strategy of \citet{DonatiBrown97} was used to construct the magnetic field topology on both stars. ZDI only measures the large-scale magnetic field as the small-scale magnetic fields cannot be recovered as the positive and negative magnetic fields within the resolution element are likely to cancel each other out. The mapping procedure involved uses the spherical harmonic expansions of the surface magnetic field, as implemented by \citet{Donati06}. The maximum spherical harmonic expansion $\ell_{max}$ = 11 was selected for HD~35296 while $\ell_{max}$ = 14 was used for HD~29615. These were the minimum values where any further increase did not produce any difference in the magnitude and topology of the magnetic field recovered. 

The magnetic maps for HD~29615 are shown in Fig. \ref{hip21632_maps} while the associated fits between the modelled data and the actual LSD profiles are shown in Fig. \ref{hip21632_V_fits}. The $\chi^2_r$ value of the magnetic models for HD~29615 was 1.25; showing that the fit accuracy was very close to reaching the noise level of the data.

\begin{figure*}
\begin{center}
\includegraphics[scale=0.60, angle=-90]{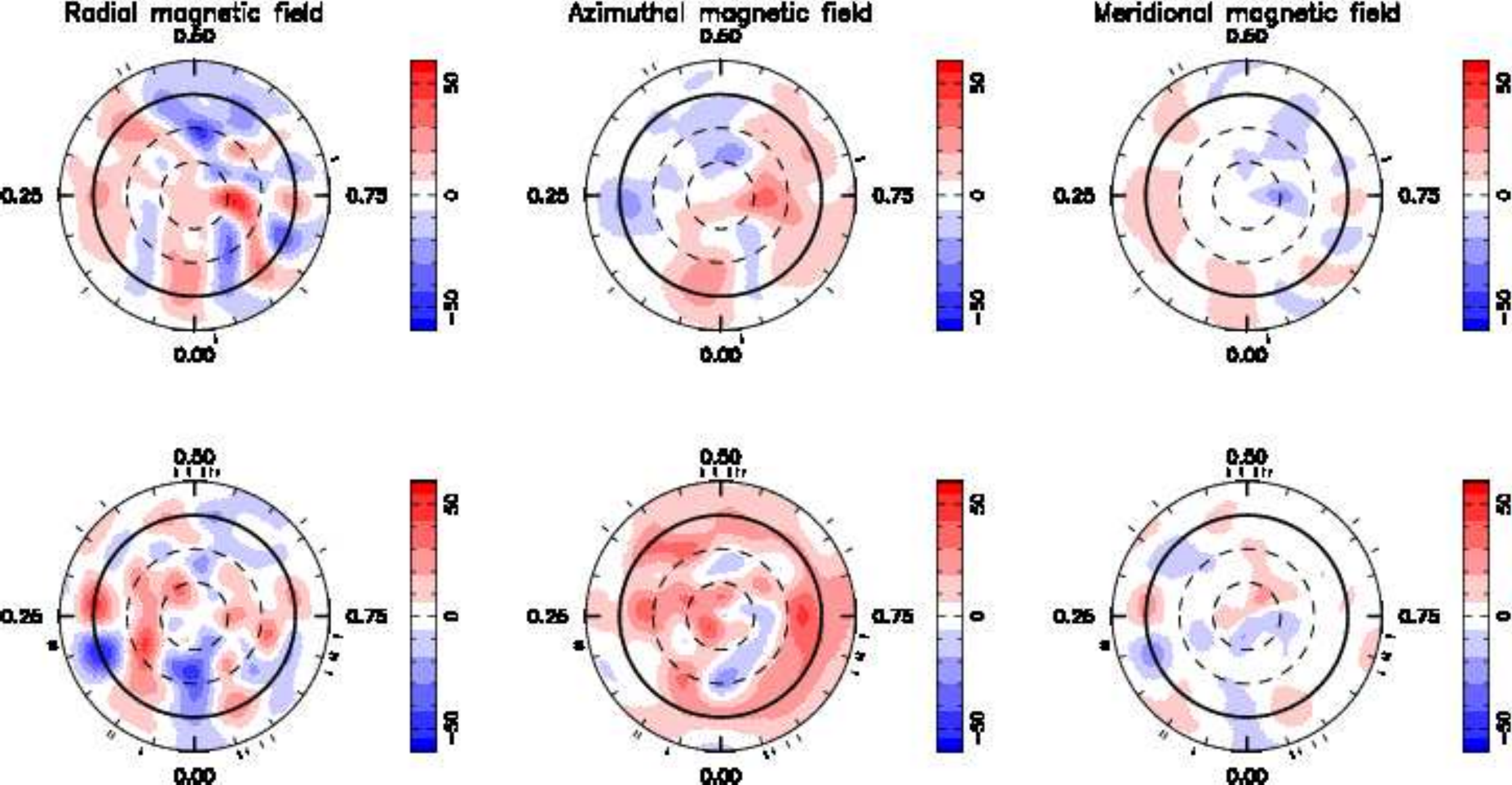}
\caption{Magnetic imaging of the large-scale field for HD~35296 from both the 2007 (top series) and 2008 (bottom series) observing runs. These maps are polar projections, as explained in Fig. \ref{hip21632_maps}. The 2007 global magnetic field was 13.4~G whereas the 2008 field was 17.9~G. The Epoch was set to MHJD = 54133.871035 for the 2007 data and MHJD = 54496.094272 for the 2008 data, with a rotational period of 3.48~d.} 
\label{HD35296_mag_maps} 
\end{center}
\end{figure*}

\begin{figure*}
\begin{center}
\includegraphics[scale=1.0, angle=0]{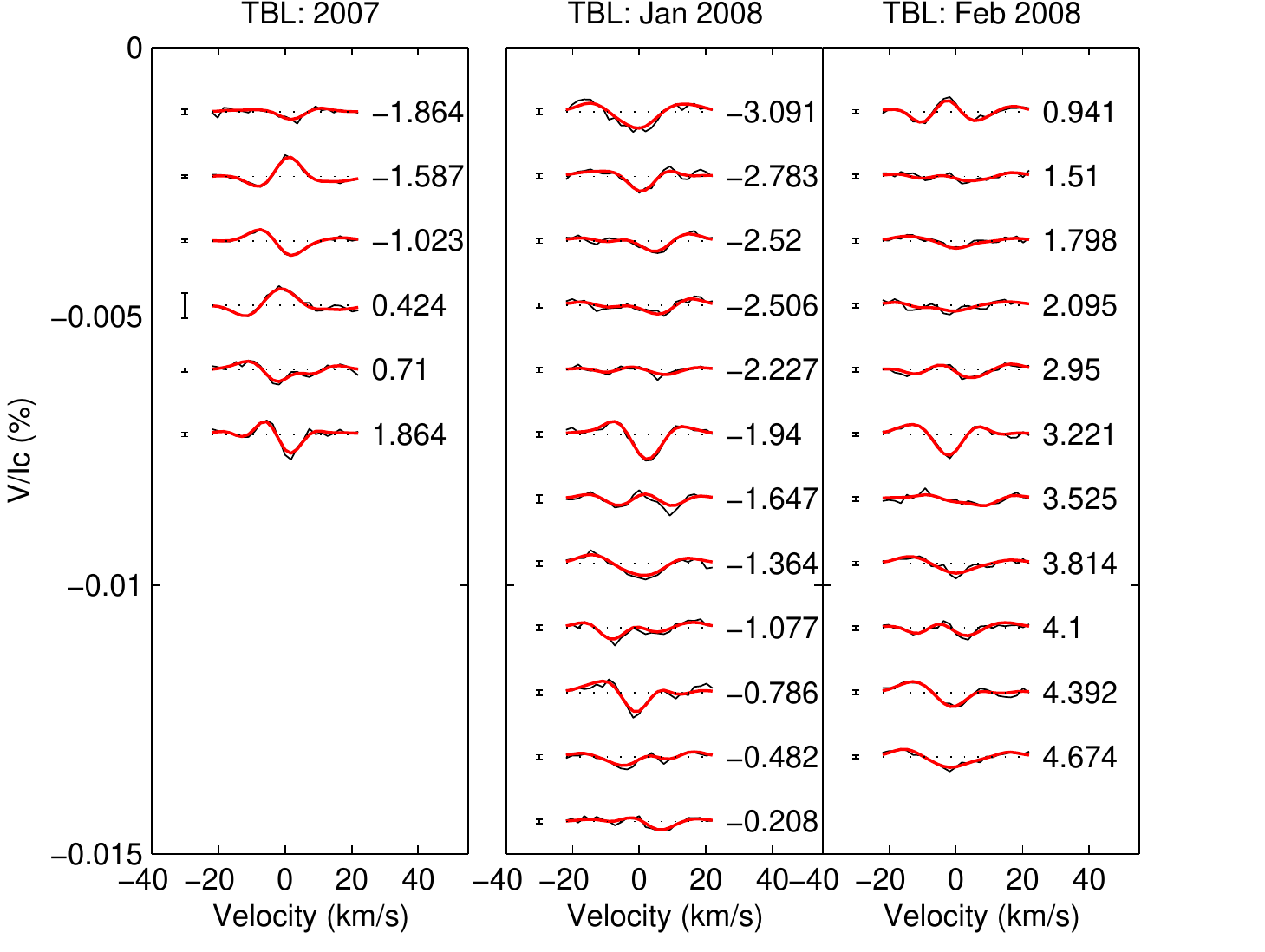}
\caption{The maximum-entropy fits to the LSD profiles for HD~35296 during 2007 (left panel) and 2008 (right panel), with differential rotation incorporated into the analysis. The observed Zeeman signatures are shown in black while the fit to the data is shown as red lines. The rotational cycle and $\pm$ 0.5 $\sigma$ error bars of each observation are shown next to each profile. Each profile has been shifted down by 0.0012 for clarity. The minimum $\chi^2_r$ value used was 1.0. The LSD profile obtained on February 02, 2007 was excluded from the mapping process due to its relatively low signal-to-noise. The rotational period used was 3.48~d.}
\label{HD35296_20078_mag_fits} 
\end{center}
\end{figure*}

The magnetic maps for HD~35296 for the 2007 and 2008 datasets are shown in Fig. \ref{HD35296_mag_maps} with the associated fits between the modelled data and the actual profiles shown in Fig. \ref{HD35296_20078_mag_fits}. The LSD profile taken on February 02, 2007 of HD~35296 was excluded from the mapping process due to the relatively poor SNR. Including these data made no difference to the resulting magnetic maps and field configuration but did make a marginal degradation of the total magnetic field strength (by 0.1~G). In each case, the minimum $\chi^2_r$ value of the magnetic models was set to 1.0. 

The full set of parameters used to produce these maps, including differential rotation (see Sect \ref{SDR}) are listed in Table \ref{parameters}. Due to the length of time between the observing runs for HD~35296, the phases for the 2007 run were calculated using an epoch of the Modified Heliocentric Julian Date (where MHJD = HJD - 2400000.0) of 54133.871035 whereas the 2008 run used an epoch of MHJD = 54496.094272. This 2008 epoch was selected as it was close to the mid-point of the observing run, but was an integer number of rotations since the mid-point of the 2007 run based on the 3.48~day period. 

\subsection{Differential rotation}
\label{SDR}

There have been a number of different methods by which differential rotation has been measured on stars. \citet{Reiners03a,Reiners03b} have used a Fourier transform method to derive the parameters for rapidly rotating  inactive F-type / early G-type stars. They produce a deconvolved line profile from the stellar spectrum and determine the ratio of the second and first zero of the resulting Fourier transform. This ratio is a measure of the magnitude of the differential rotation on the stellar surface. However, for active young stars with significant asymmetry within the line profiles, this technique is not as effective. Differential rotation of active, rapidly-rotating stars can be estimated by tracking spot features at different latitudes on the surface of the star from multiple Doppler images \citep{CollierCameron02}. Alternatively, cross-correlation techniques can be used on two independent Doppler images \citep{DonatiCameron97}. Finally, a solar-like differential rotation law, as defined in  Eqn. \ref{DR}, can be applied to measure the rotation of a number of solar-type stars \citep[e.g.][]{Petit02,Donati03b,Petit04b,Barnes05,Marsden05,Marsden06}.
\begin{equation}
	\label{DR}	
	\Omega(\theta) = \Omega_{eq} - \Delta\Omega sin^{2} \theta 
\end{equation}
where $\Omega(\theta)$ is the rotation rate at latitude $\theta$ in \rdd, $\Omega$\subs{eq} is equatorial rotation rate, and $\Delta\Omega$ is the rotational shear between the equator and the pole. This technique uses fixed information content of the spot coverage (Stokes $\it{I}$) or magnetic field strength (Stokes $\it{V}$) for the maps while systematically adjusting the differential rotation parameters, $\Omega$\subs{eq} and $\Delta\Omega$. The best fit to the data, using $\chi^{2}$ minimization techniques, is subsequently determined by fitting a paraboloid to the reduced-$\chi^{2}$ landscape thereby measuring the differential rotation parameters. It is this $\chi^{2}$ minimization technique that has been adopted for this work as it is best suited to the longer time-base data sets presented in this paper.

Using the magnetic signatures, HD~35296 has an equatorial rotational velocity, $\Omega_{eq}$, of 1.804~$\pm$~ 0.005~\rdd\ with a rotational shear, $\Delta\Omega$, of 0.22$^{+0.04}_{-0.02}$ \rdd. Using the brightness features, HD~29615 has an equatorial rotational velocity, $\Omega_{eq}$, of 2.68$_{-0.02}^{+0.06}$ \rdd\ with a rotational shear, $\Delta\Omega$, of 0.07$_{-0.03}^{+0.10}$ \rdd. Using the magnetic features, HD~29615 has an equatorial rotational velocity, $\Omega_{eq}$, of 2.74$_{-0.04}^{+0.02}$ \rdd\ with a rotational shear, $\Delta\Omega$, of 0.48$_{-0.12}^{+0.11}$ \rdd (see Table \ref{parameters}).

\begin{figure}
\begin{center}
\includegraphics[trim = 2.0cm 0.5cm 1.5cm 0.0cm,scale=0.60, angle=0]{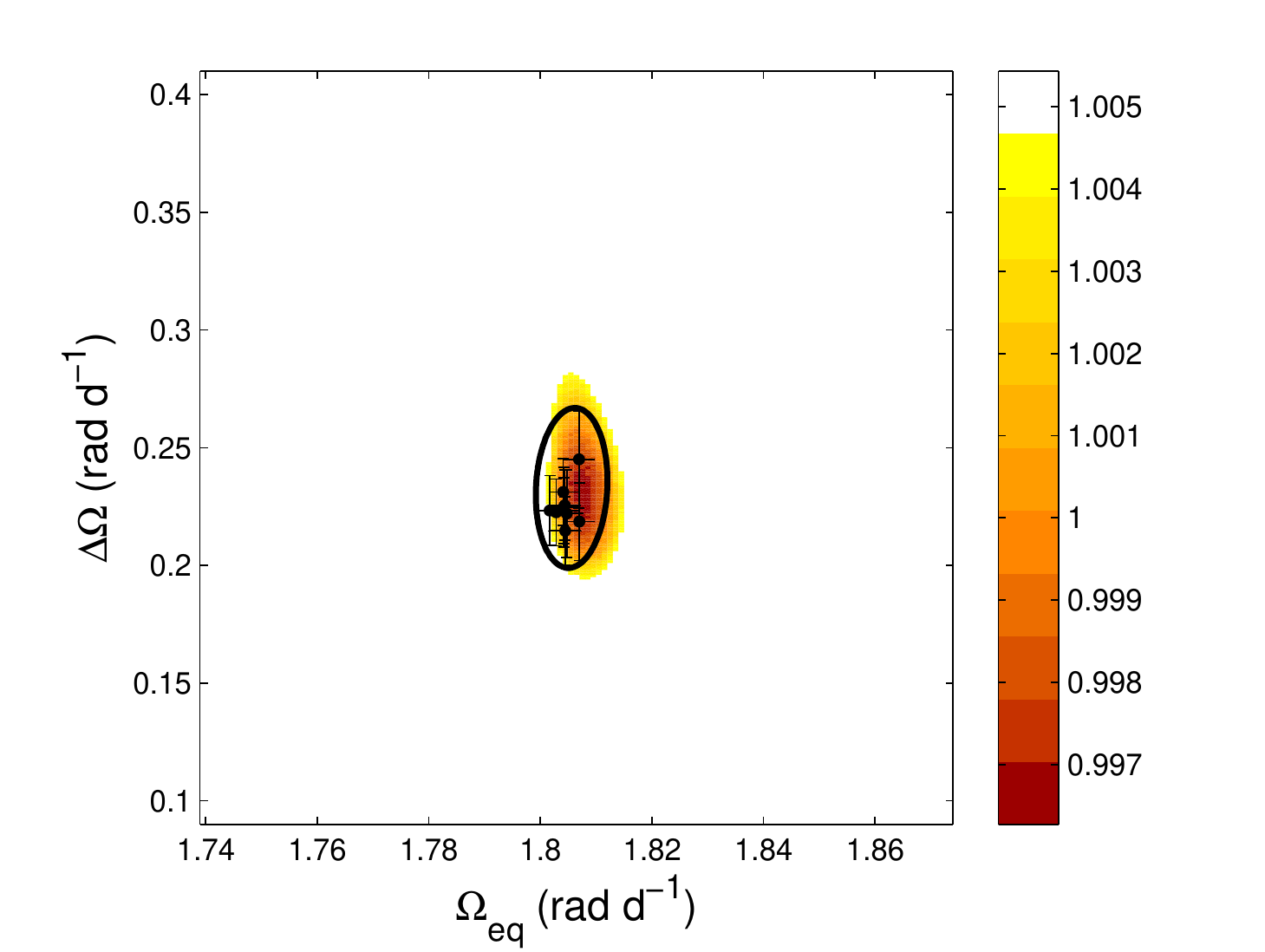}
\caption{Differential rotation using Stokes $\it{V}$: HD~35296 in 2008. The contour plot is a 1-$\sigma$ projection using the optimum set of parameters, as listed in Table \ref{parameters}. The darker regions correspond to lower reduced-$\chi^2$ values. Superimposed on this grid is the error ellipse (thick line) that encompasses a range of $\Omega_{eq}$ - $\Delta\Omega$ pairs determined by varying such parameters as \vsini, magnetic field strength and inclination. Each datapoint is the result of paraboloid fit of the reduced-$\chi^2$ landscape for each individual differential rotation value. The error bar on each datapoint are 1-$\sigma$ errors in the paraboloid fit.}
\label{HD35296_showmap} 
\end{center}
\end{figure}

\begin{figure}
\begin{center}
\includegraphics[trim = 2.0cm 0.5cm 1.5cm 0.0cm, scale=0.60, angle=0]{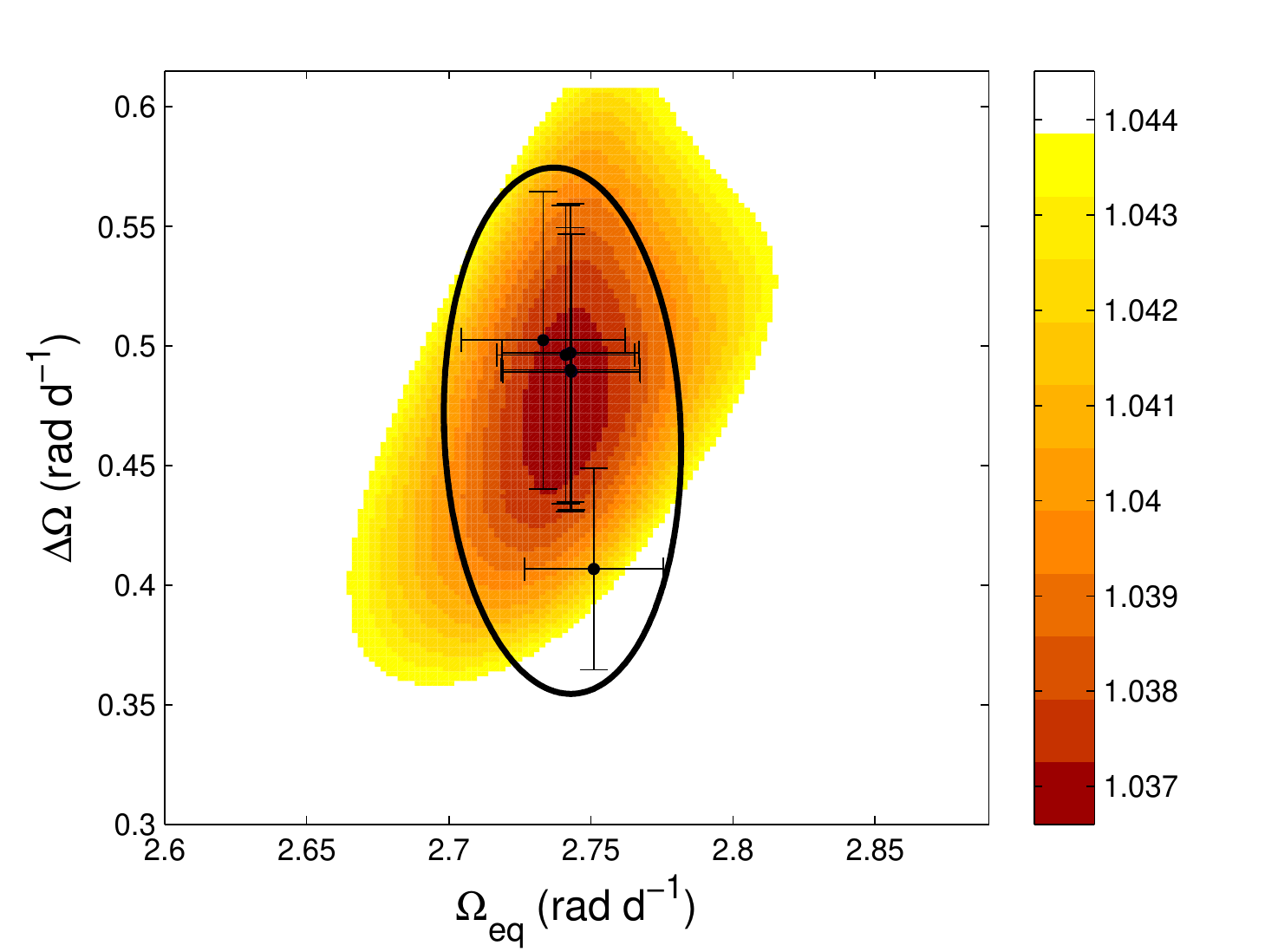}
\caption{Differential rotation using Stokes $\it{V}$: HD~29615. The contour plot is a 1-$\sigma$ projection using the optimum set of parameters, as listed in Table \ref{parameters}. Superimposed on this grid is the error ellipse (thick line) encompassing the respective $\Omega_{eq}$ - $\Delta\Omega$ pairs, as explained in Fig. \ref{HD35296_showmap}. The error bar on each datapoint are 1-$\sigma$ errors in the paraboloid fit.}
\label{HIP21632_showmap} 
\end{center}
\end{figure}

The errorsfor the differential rotation measurements were calculated by individually varying stellar parameters such as spot occupancy (for Stokes $\it{I}$: $\pm$ 10 per cent) or global magnetic field (for Stokes $\it{V}$: $\pm$ 10 per cent), inclination angle ($\pm$ 5$\degree$ for HD~35296 and +5 and -10~$\degree$ for HD~29615) and \vsini\ ($\pm$ 0.1 \kms) and determining the minimum $\Omega_{eq}$-$\Delta\Omega$ pair from each of the reduced $\chi^2$ landscapes generated. Figs. \ref{HD35296_showmap} and \ref{HIP21632_showmap} show the reduced $\chi^{2}$ landscape for the optimum set of parameters, as listed in Table \ref{parameters}. Superimposed on these are the individual $\Omega_{eq}$-$\Delta\Omega$ pairs derived from this variation of stellar parameters, with the error bars being 1-$\sigma$ errors in the paraboloid fit. The final ellipse was generated to encompass all the differential rotation values. The true error bars may be slightly larger than this due to intrinsic spot evolution during the data collection \citep[e.g.][]{Morgenthaler12}. The differential rotation parameters determined for HD~35296 for the 2008 observing run were used in the mapping process for the 2007 data as no differential rotation was able to be determined for this dataset due to the limited number of exposures.

\section{Discussion}

HD~35296 and HD~29615 are two moderately rapidly rotating young Sun-like stars that display rotational variation in their chromospheric activity. In this paper, we have determined the surface topology of both stars revealing complex magnetic fields and high levels of surface differential rotation. 

\subsection{Chromospheric activity}
\label{dis_chromo_act}
Chromospheric activity was determined using the Ca $\textsc{ii}$ H \& K, Ca $\textsc{ii}$ Infrared Triplet (IRT) and H$\alpha$ spectral lines for HD~35296 while only the H$\alpha$ line was used for HD~29615 as the spectral range of SEMPOL did not extend far enough into the respective regions to permit monitoring of the Ca $\textsc{ii}$ diagnostic lines. Table \ref{activity_indices} shows the average activity indices in 2007 and 2008 for HD~35296. The average S-index was 0.322$\pm{0.016}$ while the average H$\alpha$ index, N$_{H\alpha}$, 0.312~$\pm$~0.006 and the average Ca $\textsc{ii}$ IRT index, N$_{CaIRT}$, was 0.889$_{-0.013}^{+0.017}$. Fig. \ref{S-index} demonstrates the variable nature of the activity of HD~35296. In 2007, the chromosphere was more homogeneous with respect to phase. The large error bar at phase $\sim$ 0.15 is the result of a poor signal-to-noise observation. The amplitude of the modulation in the Ca $\textsc{ii}$ H \& K and H$\alpha$ indices is less than that observed in 2008. This would indicate that HD~35296 is more active during the 2008 epoch. In addition, the error bars on each of the data points do not overlap indicating that the chromosphere not only varied as a function of rotation, but also from one rotation to the next. This is clearly observed in all three chromospheric diagnostic lines. The conclusion is that the chromosphere of HD~35296 is highly variable over a one month interval of approximately eight stellar rotations.

\begin{figure}
\begin{center}
\includegraphics[trim = 2.0cm 0.0cm 1.5cm 0.0cm,scale=0.34, angle=0]{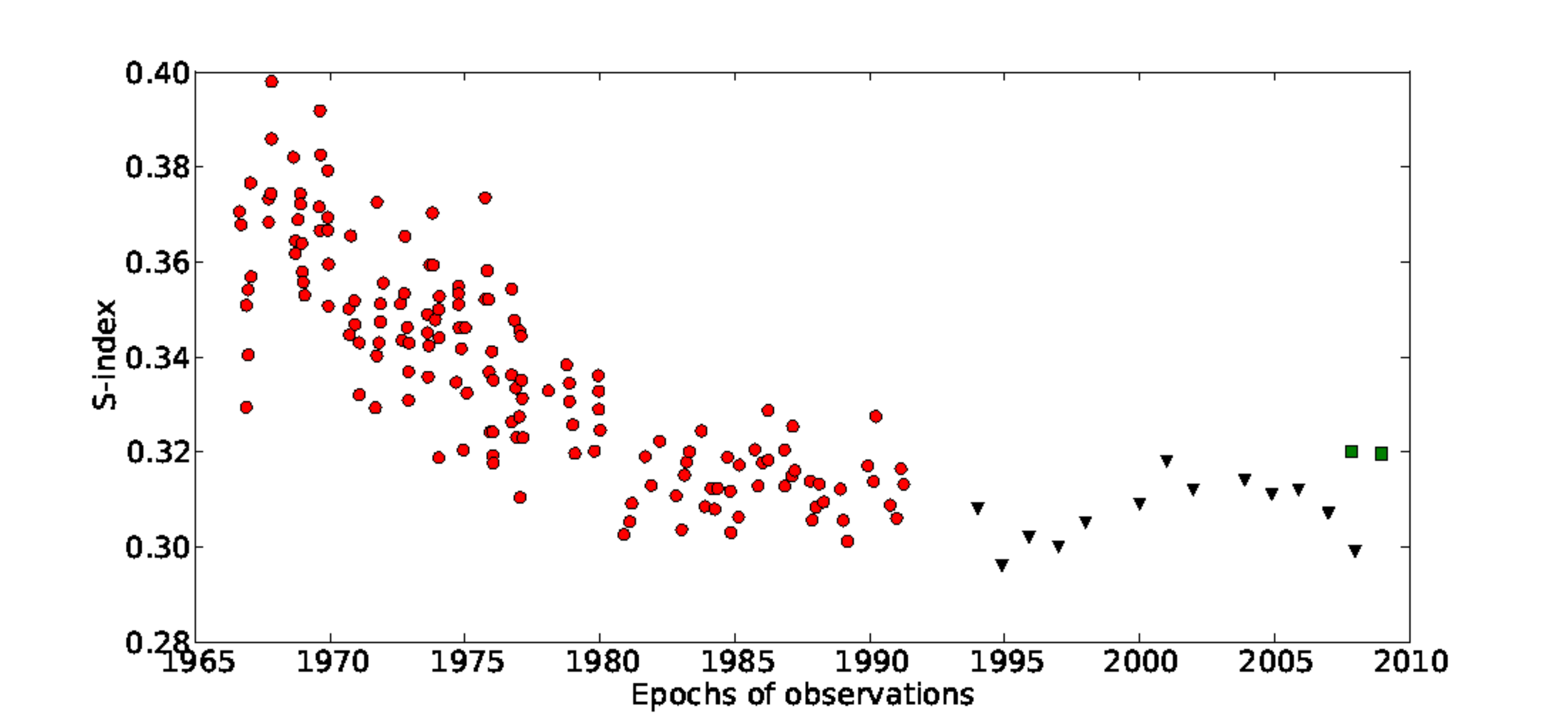}
\caption[]{The long-term Ca $\textsc{ii}$ H \& K S-index data for HD~35296. The Mount Wilson data taken from 1966 to 1991, shown as \tikz\draw[black,fill=red] (0,0) circle (0.5ex); \citep{Baliunas95} while the Lowell data taken from 1994 to 2008, shown as $\blacktriangledown$ \citep{Hall09}. Additionally, the two values shown as \tikz\draw [black,fill=green] (0.3,0.3) rectangle (0.2,0.2); were determined by this work.}
\label{HD35296_LongTerm_Sindex} 
\end{center}
\end{figure}

S-index measurements  are also available from  \citet{Baliunas95}, where data was obtained between 1966 to 1991. The average S-index from the Baliunas sample is 0.33~$\pm$~0.02 and the {\it Log R'$_{HK}$} is -4.36~$\pm$~0.04. Mount Wilson data is not available from 1991 onwards but fortunately Lowell observatory started a long-term Ca {\sc ii} H \& K monitoring programme where HD~35296 was observed from 1994 to 2008 \citep{Hall09}. All available S-index values scaled to the Mount Wilson system are shown in Fig. \ref{HD35296_LongTerm_Sindex}. HD~35296 exhibits a gradual decrease in its S-index from 1966 to 1980 and shows a relatively flat trend from 1980 onwards.  Interestingly during the low activity period the S-index does exhibit small variations. The long-term variations also indicate a probable long cyclic period which was also determined by \citet{Baliunas95}. No error estimates are available for the Lowell catalogue.

The average H$\alpha$ index for HD~29615 is 0.373~$\pm$~0.008. As indicated in Sect. \ref{chromo_activity}, the H$\alpha$ spectral line is often associated with plages and prominences \citep[e.g.][]{Thatcher93,Montes04}. There was evidence of the existence of plage-like behaviour on HD~29615 when considering the H$\alpha$ emission as shown in Fig. \ref{hip21632_halpha} (top panel). If longitude increases with phase with 0$\degree$ longitude corresponding to 0.0 phase, then this H$\alpha$ emission reaches a maximum at longitude $\sim$220$\degree$. There appears to be enhanced magnetic features in the radial magnetic field trailing slightly, at approximately 180$\degree$, as shown in Fig. \ref{hip21632_halpha}. The three lower panels in Fig. \ref{hip21632_halpha} show the fractional magnetic field strength as a function of longitude around the star for the radial, azimuthal and meridional magnetic field components. These were calculated using Eqn. \ref{Long}.

\begin{equation}
	\label{Long}	
	F(\varphi) =  {B(\varphi)d\varphi}
\end{equation}
where F($\varphi$) is the fractional magnetic field strength at longitude $\varphi$, B($\varphi$) is the average magnetic field strength at longitude $\varphi$, and $\it{d}$$\varphi$ is the width of each latitude ring. These enhanced features can also be seen in the resulting map in Fig. \ref{hip21632_maps} (top right panel). We can speculate that the enhanced positive radial magnetic field is contributing to the activity in the mid-level chromosphere, perhaps with the production of plages or even prominences at mid-latitude, very similar in location to those features observed on the Sun.


\subsection{Brightness and Magnetic Maps}
\label{BrightnessMagnetic}
The results presented in this paper show that HD~35296 and HD~29615 are young, Sun-like stars whose moderately rapid rotation has led to very complex large-scale surface magnetic field topology. As discussed in Sect. \ref{ImageReco}, brightness images were not produced for HD~35296 due to its relatively low \vsini\ with the LSD intensity profiles not being sufficiently deformed by any surface features present to recover reliable spot information using DI. The brightness map for HD~29615, in Fig. \ref{hip21632_maps} (top left panel), shows a dominant polar spot. This polar spot is slightly off-centred with features extending to $\sim$60$\degree$ latitude for phases between, $\phi$, 0.0 to 0.5, but not evident for $\phi$ = 0.5 to 1.0. While the polar spot is very prominent, there appears limited spot coverage at mid- to lower-latitudes with a small number of features observed at $\phi$ $\sim$ 0, 0.25-0.3 and 0.55-0.65. The features at $\phi$ $\sim$ 0 and $\sim$ 0.25 may not be reliable as the dynamic spectrum shown in Fig. \ref{hip21632_residuals} displays a dark diagonal band moving from the blue wing to the red wing. This indicates that the modelled data may have incorporated spots at this phase that were not present in the original data hence could not accurately constrain the latitude of these two features. \citet{Unruh97} note that spectroscopy extracts the higher latitude features but cannot discriminate lower latitude features at latitudes $\la$ 30$\degree$. However, the paucity of spots at mid- to lower-latitudes is dissimilar to other more rapidly rotating solar-type stars like He~699 (V532~Per, SpType: G2-3V) \citep{Jeffers02}, HD~106506 (SpType: G1V) \citep{Waite11a} and HD~141943 (SpType: G2) \citep{Marsden11a} that appear to have spots covering all latitudes and phases.  


The magnetic maps for HD~29615, in Fig. \ref{hip21632_maps}, show azimuthal, meridional and radial magnetic field structures. The azimuthal magnetic field is strongly positive although not a complete ring of field, unlike that observed on other solar-type stars such as HD~106506 \citep{Waite11a} and HD~141943 \citep{Marsden11a}. The radial field is also strongly positive at higher latitudes, from approximately +30$\degree$ to the pole. The maps for HD~35296 from 2007 and 2008 are shown in Fig. \ref{HD35296_mag_maps}. Like HD~29615, HD~35296 displays a predominantly positive azimuthal magnetic field in both 2007 and 2008 while the radial and meridional fields show mixed polarity across the surface. 

When using Stokes $\it{V}$ without Stokes $\it{I,Q,U}$ there is crosstalk between the three magnetic field configurations, although the azimuthal field is the most robust with limited crosstalk. At high latitudes, ZDI can recover all three field components but as lower latitudes, in particular for stars with inclination angles greater than 50$\degree$, the crosstalk is from the meridional to radial field component \citep{DonatiBrown97}. This crosstalk appears stronger for HD~35296 when compared with HD~29615. This could be due to either a weaker magnetic field or fewer phases for each rotation for HD 35296 thereby affecting the code's ability to distinguish between the radial and meridional field components, particularly at lower latitudes. \citet{Rosen12} have observed when using all four Stokes $\it{IQUV}$ parameters, the crosstalk between the components is significantly reduced. However, the linear polarization state (Stokes $\it{Q}$ and $\it{U}$ parameters) are significantly weaker than the Stokes $\it{V}$ and are extremely difficult to observe on active cools stars such as HD~29615 and HD~35296 \citep[e.g.][]{Wade00,Kochukhov04}.

\subsection{Latitude dependence of the magnetic fields}
\label{LatitudeDependence}

\begin{figure*}
\begin{center}
\includegraphics[scale=1.1, angle=0]{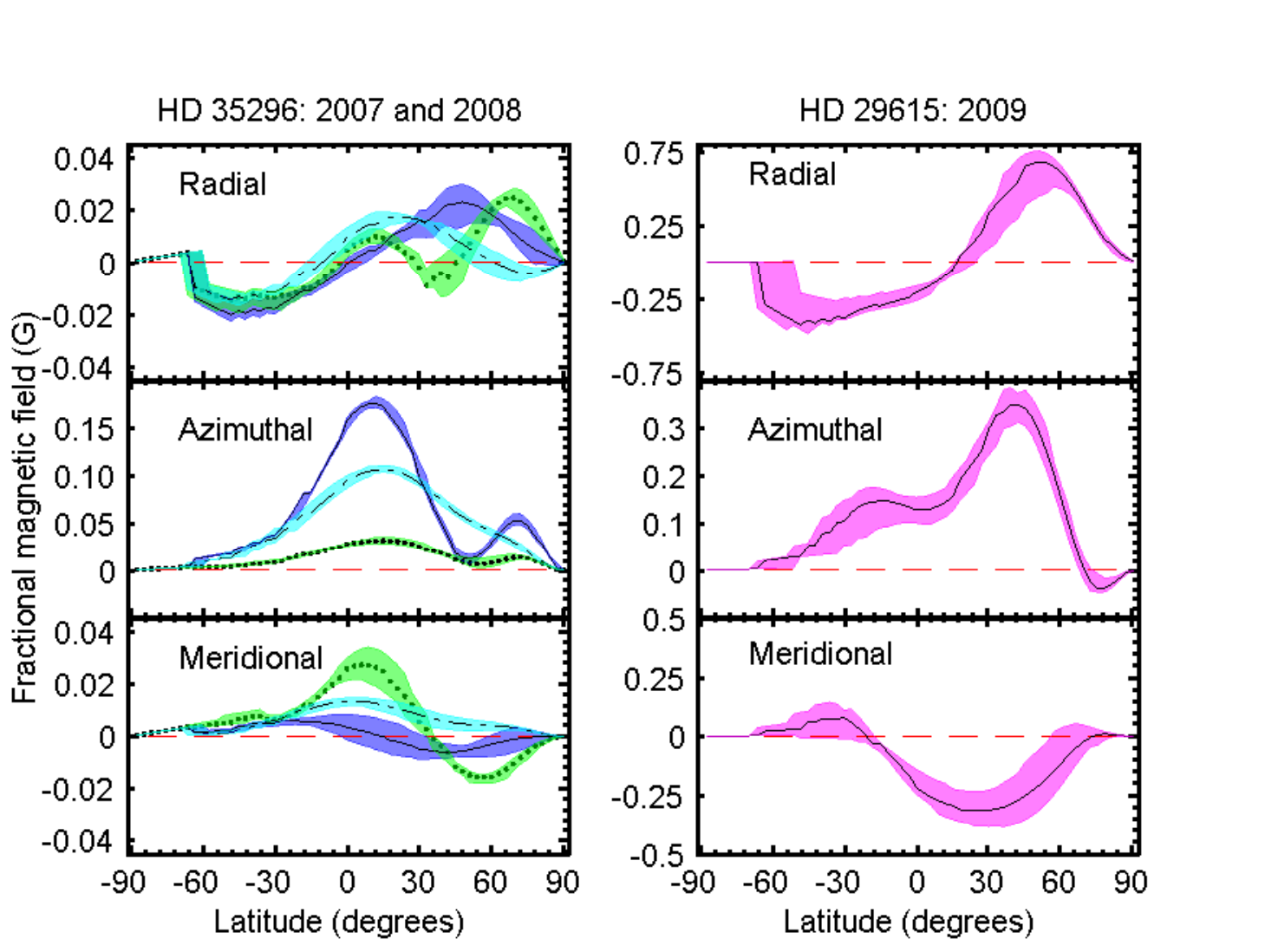}
\caption{Variation of the respective magnetic field orientations for HD~35296 (left series of panels) and HD~29615 (right series of panels). The shaded regions show the variation in the strength at each latitude point as determined by varying \vsini, inclination angle, $\Omega_{eq}$ and $\Delta\Omega$ within the limits of their respective accuracy (see Sect. \ref{SDR}). Considering HD~35296, the solid line with the dark blue shading is for 2008 data while the dotted line with the green shading is the 2007 data. The dot-dash line with cyan shading represents the 2008 data with reduced number of profiles approximately matching the number of profiles and phase of the 2007 data. The period used for HD~35296 was 3.48~d while the period used for HD~29615 was 2.34~d.}
\label{B_V_Latitude} 
\end{center}
\end{figure*}

The fractional magnetic field strength as a function of latitude for each orientation for HD~35296 and HD~29615 was determined using Eqn. \ref{V_spottedness}: 

\begin{equation}
	\label{V_spottedness}	
	F(\theta) =  \frac{B(\theta)cos(\theta)d\theta}{2}
\end{equation}
where F($\theta$) is the fractional magnetic field strength at latitude $\theta$, B($\theta$) is the average magnetic field strength at latitude $\theta$, and $\it{d}$$\theta$ is the width of each latitude ring. The results are shown in Fig. \ref{B_V_Latitude}. In order to determine the robustness of each measurement, the \vsini, v$_{rad}$,inclination angle, $\Omega_{eq}$ and $\Delta\Omega$ were varied, placing limits on the fractional magnetic field at each latitude for each of the orientations. These measured variations are shown by the shaded regions while the values for the optimum parameter set are shown as a solid line for the 2007 data and a dotted line for the 2008 data. HD~35296 displayed both positive and negative radial fields in 2007 but predominantly positive field in 2008 whereas the azimuthal field remains a positive field for both epochs. Conversely, the radial field, at mid-latitudes, appears to have undergone a reversal of polarity from 2007 to 2008. This may be real or an artefact due to fewer profiles being used in the reconstruction. 

To test this, we reduced the number of profiles in the 2008 dataset, to match the same number of profiles and phase as that obtained in 2007. This had an impact on the strength of the global magnetic field (17.9~G reduces to 11.9~G) and the latitudinal distribution of the respective fields. The azimuthal field remained strongly positive, as shown in Fig. \ref{B_V_Latitude} (dot-dash line with cyan colour). However, the radial field no longer shows a strong positive peak at $\sim$ 70$\degree$ latitude when compared with using the full number of profiles, although it does remain positive at these mid-latitudes. As a further test, maps were produced from the first two rotations, middle four rotations and final two rotations. This demonstrated the robust nature of the azimuthal magnetic field, although fewer profiles meant that the imaging code could not adequately recover the latitude of each feature. Nevertheless, the mid-latitude reversal in the radial field from 2007 to 2008 may well be real. The evidence is tantalising but more data would be required before any definitive conclusion can be drawn regarding any changes in the mid-latitude field polarity of HD~35296.

Comparing HD~35296 with HD~29615, it appears that the dominant azimuthal fields are restricted to the equatorial regions for HD~35296 whereas the similar fields dominate at higher latitudes for HD~29615. Both HD~35296 and HD~29615 have similar latitudinal distributions in the meridional and radial fields. Thus both of these young, Sun-like stars with similar masses and \vsini\ have similar magnetic field configurations.

\subsection{Magnetic field configurations}

Large-scale magnetic fields on the Sun are considered to arise from the interplay between the poloidal and toroidal components in what is commonly referred to as the $\alpha-\Omega$ dynamo \citep{Babcock61}. In the Sun, the toroidal component is understood to be confined to the interface between the radiative core and the convection zone \citep[e.g.][]{MacGregor97,Charbonneau97}. However, in solar-type stars, the toroidal component may manifest itself in the form of strong azimuthal magnetic fields on, or near, the surface of the star \citep[e.g.][]{Donati03a,Petit04b}. Additionally, \citet{Petit08} infer that a rotation period lower than $\sim$12~d is necessary for the toroidal magnetic energy to dominate over the poloidal component. Both of these moderately rapidly rotating stars appear to have a large azimuthal component. These results are consistent with observations of more rapidly rotating stars exhibiting similarly strong azimuthal magnetic fields such as AB~Doradus (SpType: K0V), LQ~Hydrae (SpType: K0V) and HR~1099 (SpType: K2:Vnk) \citep{Donati03a}, HD~171488 (SpType: G2V) \citep{Marsden06,Jeffers08,Jeffers11}, HD~141943 (SpType: G2) \citep{Marsden11b} and HD~106506 (SpType: G1V) \citep{Waite11a}. Observations of these strong surface azimuthal magnetic fields on active stars have been interpreted by \citet{Donati03a} as a result of the underlying dynamo processes being capable of generating fields directly in the subsurface region \citep[e.g.][]{Dikpati02} and could be distributed throughout the whole convection zone \citep[e.g.][]{Lanza98}. 

\begin{table*}
\begin{center}
\caption{Magnetic quantities derived from the set of magnetic maps for HD~35296 and HD~29615. Both the poloidal and toroidal components are listed along with the relative weightings in each geometry. The errors are variation bars, as described in Sect. \ref{SDR}.} 
\label{magpot}
\begin{tabular}{lcccccccc}
\hline
 $\it{B_{mean}}$ & geometry   & Energy$^1$  & dipole$^2$ & quad.$^2$ & oct.$^2$ & higher$^2$      & axi.$^2$ & axi.$^1$ \\
     (G)         &            & (\%)  	  & (\%)       & (\%)      & (\%)     & order (\%) 	& (\%) 	   & (\%)	 \\
\hline
\multicolumn{9}{l}{HD~35296: 2007 data}            		\\ 	  
 13.4$^{+0.8}_{-0.4}$	& poloidal   & 82$^{+2}_{-1}$   &  20$^{+1}_{-4}$    &  7$^{+1}_{-1}$   & 12$^{+0}_{-2}$   &  61$^{+4}_{-1}$   &  7$^{+7}_{-2}$ & 6$^{+5}_{-2}$		\\	
	& toroidal   & 18$^{+1}_{2}$ &  33$^{+1}_{-11}$	 &  14$^{+2}_{-1}$ &  5$^{+3}_{-0}$   &  48$^{+9}_{-2}$ & 72$^{+1}_{-10}$ & 13$^{+1}_{-2}$		\\
\multicolumn{9}{l}{HD~35296: 2008 data}            	\\ 		
 19.7$^{+0.7}_{-1.3}$	& poloidal   & 50$^{+2}_{-3}$   &   7$^{+1}_{-2}$    & 12$^{+1}_{-1}$  & 8$^{+2}_{-0}$   &  73$^{+2}_{-4}$ & 13$^{+3}_{-3}$ & 7$^{+1}_{-2}$		\\
	& toroidal   & 50$^{+3}_{-2}$   &  53$^{+6}_{-1}$    & 6$^{+4}_{-1}$   & 3$^{+1}_{-1}$   &  38$^{+1}_{-9}$ & 76$^{+5}_{-0}$ & 38$^{+4}_{-1}$		\\
\multicolumn{9}{l}{HD~35296: 2008 data with reduced number of profiles to match the 2007 dataset}           		\\ 
11.9$^{+0.0}_{-1.0}$	& poloidal & 48$^{+0}_{-4}$  & 25$^{+2}_{-4}$     & 16$^{+9}_{-0}$   &  16$^{+0}_{-6}$  & 43$^{+2}_{-3}$    & 13$^{+0}_{-6}$   & 6$^{+0}_{-3}$	\\ 
	& toroidal & 52$^{+4}_{-0}$     & 71$^{+8}_{-2}$     & 12$^{+4}_{-4}$   &   4$^{+1}_{-2}$  & 13$^{+0}_{-3}$   & 87$^{+4}_{-0}$  & 46$^{+4}_{-0}$     	\\
\hline
\multicolumn{9}{l}{HD~29615: 2009 data}	             \\
81.6$^{+20}_{-15}$      & poloidal  & 75$^{+6}_{-5}$   &  58$^{+7}_{-13}$    & 10$^{+3}_{-2}$   & 13$^{+3}_{-5}$   &  19$^{+19}_{-0}$  & 66$^{+2}_{-12}$ &	49$^{+6}_{-10}$ \\
	& toroidal  & 25$^{+5}_{0}$	 &  34$^{+0}_{-8}$	 & 18$^{+6}_{-3}$   &	10$^{+3}_{-1}$  &  38$^{+9}_{-3}$	 & 67$^{+3}_{-7}$ &	17$^{+5}_{-2}$ \\
\hline
\end{tabular}
\end{center}
$^1$ this is a fraction (in \%) of the total magnetic energy available. \\
$^2$ this is a fraction (in \%) of the respective poloidal or toroidal field energy. \\
Listed also is the fraction of the poloidal or toroidal magnetic energy in the dipolar ($\ell$ = 1), quadrupolar ($\ell$ = 2), octupolar ($\ell$~=~3) and higher order ($\ell$ $\geq$ 4) components as well as the fraction of energy stored in the axisymmetric component (m = 0).
\end{table*}

The magnetic field on HD~35296 was predominantly poloidal in 2007, with $\sim$82$_{-1}^{+2}$ per cent of the total magnetic energy. By 2008, the field had reorganised itself to a $\sim$50:50 per cent poloidal-toroidal configuration. This is shown in Table \ref{magpot} and in Fig. \ref{HD35296_mag_maps}. Again there are two plausible reasons for this. One is that the magnetic topologies had significantly changed. Alternatively, simply using fewer profiles may be responsible for the observed changes. Reducing the number of profiles, as explained in Sect. \ref{LatitudeDependence}, had only a minor effect on the balance of the poloidal-toroidal field configuration in 2008. This is shown in Table \ref{magpot}, where the full dataset produced a poloidal-toroidal ratio of 50:50 per cent while the reduced dataset produced a ratio of 48 to 52 per cent. Higher cadence data would be required to support the conclusion that the global magnetic field topology had significantly evolved during the course of one year on HD~35296. The second star, HD~29615, was observed to be strongly poloidal with 75$_{-5}^{+6}$ per cent of the magnetic energy being held in this configuration. Of this, 58 per cent of the poloidal component is dipolar ($\ell$ = 1). This is similar to other stars such as HD 76151 (SpType: G3V) \citep{Petit08}, $\tau$ Bootis (SpType: F6IV) \citep{Fares09} and $\xi$ Bootis (SpType: G8V) \citep{Morgenthaler12} which are also strongly poloidal with complex fields with predominantly dipolar fields. Table \ref{magpot} gives a full listing of the various components for both HD~35296 and HD~29615.

\subsection{Differential rotation}

Differential rotation was measured for HD~35296 and HD~29615. The differential rotation of HD~35296, using Stokes $\it{V}$, was measured for the 2008 data using the $\chi^2$ minimization technique, as described in Sect.~\ref{SDR}. Using the magnetic signatures, HD~35296 has an equatorial rotational velocity, $\Omega_{eq}$, of 1.804~$\pm$~0.005~\rdd\ with rotational shear, $\Delta\Omega$, of 0.22$^{+0.04}_{-0.02}$ \rdd. Differential rotation was not observed on the 2007 data due to the limited number of profiles obtained. This $\Delta\Omega$ value is consistent with similar observations of other F-type stars by \citet{Reiners06} and \citet{Ammler12}; however, \citet{Reiners06} found no evidence of differential rotation using the Fourier transform method of line profile analysis on HD~35296. Another star that exhibits this discrepancy is the pre-main-sequence binary star HD~155555 (SpType = G5IV+K0IV) when \citet{Dunstone08} measured differential rotation on both components while \citet{Ammler12} could not detect differential rotation. \citet{Reiners06} conducted a direct comparison between the Fourier transform method and Doppler imaging using HD~307938 (R58) in IC~2602. \citet{Marsden05} used DI to measure a shear of $\Delta\Omega$ = 0.025 $\pm$ 0.015 \rdd. Using the Fourier transform method, the threshold for solid-body rotation, $\it{q_{2}/q_{1}}$, is 1.76 where $\it{q_{1}}$ and $\it{q_{2}}$ are the first two zeros of the line profile's Fourier transform. \citet{Reiners06} argued that for a star with a large polar spot and small shear (such as R58), the spot has more influence on $\it{q_{2}/q_{1}}$ than small deviations from solid-body rotation. This does not answer the question why HD~35296's differential rotation was measured using ZDI but not Fourier transform method. \citet{Reiners06} measured  $\it{q_{2}/q_{1}}$ = 1.75 (for HD~35296), only marginally less than solid-body rotation, yet this work measured a rotational shear, $\Delta\Omega$, of 0.22$^{+0.04}_{-0.02}$~\rdd. One would expect this to give a $\it{q_{2}/q_{1}}$ $\sim$1.50 to 1.60. This disparity in measured differential rotation could be real, or just reflects the fact that this work is based on Stokes $\it{V}$ whereas the value of \citet{Reiners06} was derived from Stokes $\it{I}$ data which often gives lower values, as shown by the results for HD~29615. Alternatively, the Stokes $\it{I}$ LSD profiles of HD~35296 were not sufficiently deformed by spots to produce a detailed map whereas their presence may have affected the measurement in the Fourier domain. 

\begin{figure}
\begin{center}
\includegraphics[trim = 5cm 0cm 5cm 0cm, scale=0.90, angle=0]{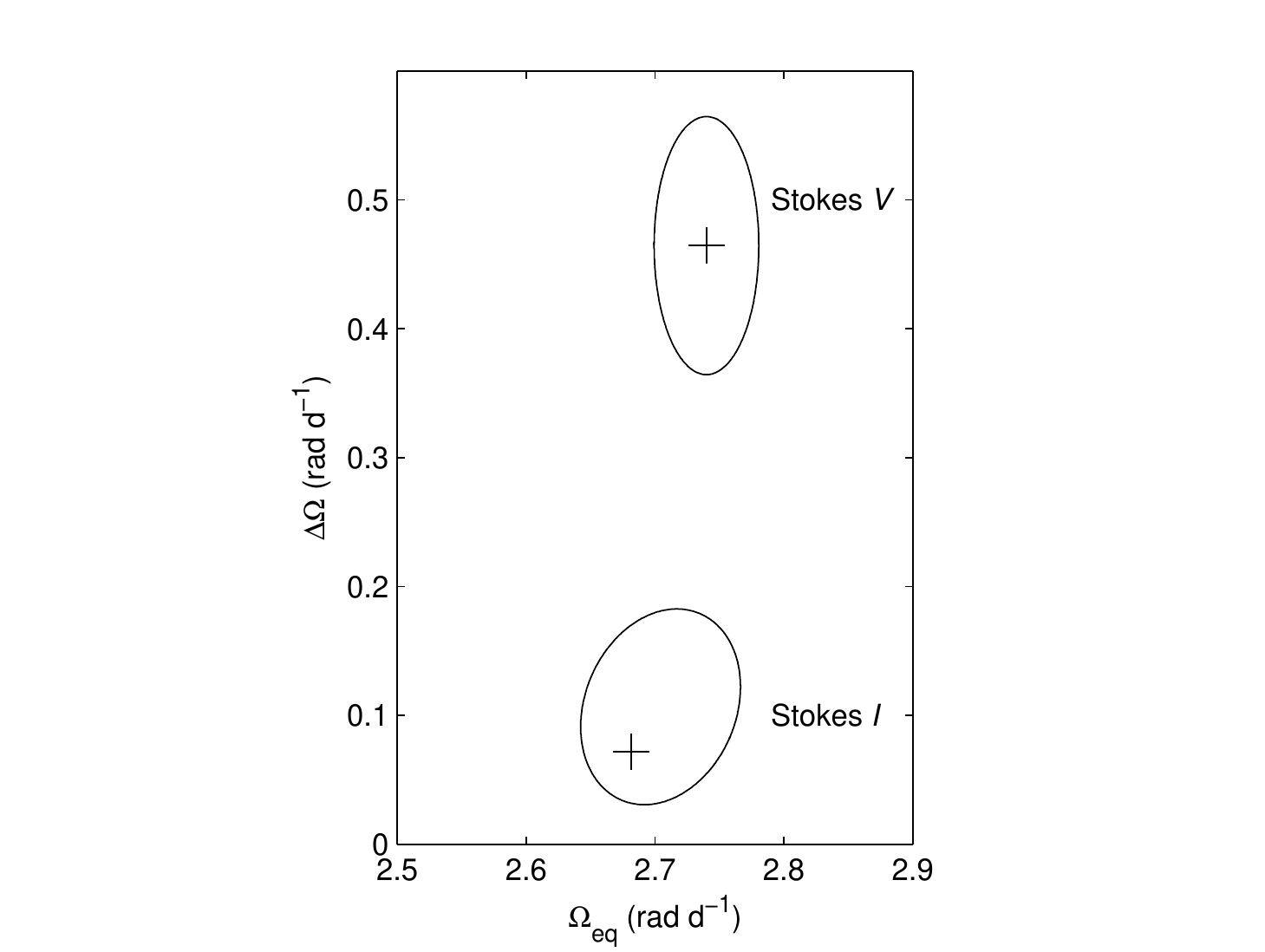}
\caption{Differential rotation using Stokes $\it{I}$ and $\it{V}$ for HD~29615, including the associated variation ellipses (as explained in Sect. \ref{SDR}). The brightness (spot) features produce an equatorial rotational velocity, $\Omega_{eq}$ of 2.68$_{-0.02}^{+0.06}$ \rdd\ with rotational shear, $\Delta\Omega$ of 0.07$_{-0.03}^{+0.10}$ \rdd. Using the magnetic features observed, HD~29615 has an equatorial rotational velocity, $\Omega_{eq}$ of 2.74 $_{-0.04}^{+0.02}$ \rdd\ with rotational shear, $\Delta\Omega$ of 0.48$_{-0.12}^{+0.11}$ \rdd.}
\label{HIP21632_errorellipse} 
\end{center}
\end{figure}

The differential rotation measurement for HD~29615, using the magnetic features indicate an equatorial rotational rate $\Omega_{eq}$ = 2.74$_{-0.04}^{+0.02}$ \rdd\ and shear $\Delta\Omega$ = 0.48$_{-0.12}^{+0.11}$ \rdd. This rotational shear is relatively large although \citet{Reinhold13} have measured larger rotational shear on active $\it{KEPLER}$ stars with temperatures commensurate with HD~29615 (see Fig. 15 of that work). In contrast, the spot features provide a different value, namely $\Omega_{eq}$ = 2.68$_{-0.02}^{+0.06}$ \rdd\ and $\Delta\Omega$ of 0.07$_{-0.03}^{+0.10}$ \rdd (see Fig. \ref{HIP21632_errorellipse}). \citet{Donati03b} noted that in some early K-dwarf stars, the level of differential rotation measured from brightness features is usually lower when compared with the levels measured when using magnetic features. Their interpretation of this variation is that the brightness features and the magnetic features are anchored at different depths within the convection zone of the star. This could also be the case here, though the difference in $\Delta\Omega$ that we measure for HD~29615 is more extreme than previously observed for other stars. An alternative explanation may be offered by the work of \citet{Korhonen11}. Using dynamo calculations, they suggest that large starspots do not necessarily follow the actual differential rotation of the star, but have more solid-body like behaviour, whereas the true surface differential rotation is only recovered if small magnetic features are added to the simulations. In addition, the paucity of low level spot features could affect the differential rotation measurements. The differing measurements of differential rotation should be treated with caution until further Stokes $\it{I}$ and $\it{V}$ data are obtained for this star.


\subsection{Surface differential rotation, rotation rate and convection zone depth}

\citet{Barnes05} showed a relationship between differential rotation ($\Delta\Omega$) and temperature, with increasing shear as a result of increasing temperature. Recent studies of G-dwarf stars such as HD~171488 \citep{Marsden06,Jeffers08,Jeffers11} and HD~141943 \citep{Marsden11b} show significant levels of differential rotation beyond those observed by \citet{Barnes05}. HD~29615, like HD~171488, has extreme levels of rotational shear with a $\Delta\Omega$ = 0.48$_{-0.12}^{+0.11}$ \rdd.  HD~171488 has a 1.33~d rotation period with a \vsinis = 38 \kms\ whereas HD~29615 has a rotational period of 2.34~d with a \vsinis = 19.6 \kms. Given that both HD~171488 and HD~29615 have similar photospheric temperatures and age, it appears that rotation rate has limited effect on rotational shear, thereby supporting the observations by \citet{Barnes05}. 

\citet{Kuker11}, using theoretical models, demonstrated that the extreme surface shear of stars such as HD~171488 (and by implications HD~29615) can only be explained with a shallow convection zone. HD~171488 has a convection zone depth (CZD) of 0.206~$R_{\star}$ (0.233~$R_{\odot}$) \citep{Jeffers11} yet the CZD for HD~29615 was estimated to be 0.252 $\pm$ 0.011~$R_{\star}$, as determined from the stellar evolution models of \citet{Siess00}. The estimate for HD~29615 was based on the V-I colour index \citep{Van_Leeuwen07} and the absolute magnitude, as determined from the maximum visual magnitude measured by the $\it{HIPPARCOS}$ space mission. \citet{ONeal96} concluded that on some heavily spotted stars, the observed maximum V magnitude underestimates the brightness of the unspotted star by $\sim$ 0.3-0.4~mag. For example, HD~106506 \citep{Waite11a} had a visual magnitude of $\sim$8.54 but the unspotted magnitude was determined to be 8.38 (using the imaging code). As there was limited knowledge of the fractional spottedness of many of the stars that had their differential rotation already determined by other authors it was decided to use the maximum visual magnitude listed in the $\it{HIPPARCOS}$ database with the caveat that this is possibly leads to an overestimate of the true CZD. This measurement for HD~29615 places additional constraints on the conclusion of the theoretical work of \citet{Kuker11} that extreme levels of differential rotation require a shallow convection zone.

\section{Conclusions} 

From the results presented in this paper, HD~35296 and HD~29615 are both Sun-like stars whose moderately rapid rotation has led to very complex surface magnetic fields that exhibit high levels of chromospheric activity and surface differential rotation. This variation in the chromospheric activity, with rotational phase, was evidenced by modulation of the Ca \textsc{ii} H \& K, Ca \textsc{ii} Infrared Triplet and H$\alpha$ spectral lines (HD~35296) and H$\alpha$ emission (HD~29615). High levels of differential rotation were measured. The differential rotation on HD~29615 showed a significant discrepancy in shear values between spot and magnetic features. This is an extreme example of a variation also observed for other lower-mass stars. There are indications that the magnetic field of HD~35296 appears to have undergone a minor reorganisation of its field from 2007 to 2008 thereby demonstrating the evolving nature of the magnetic topology on this star. We conclude that the dynamo operating on both of these stars is similar to that of other more active, rapidly rotating stars and is most likely a distributed dynamo operating throughout the convection zone.

\section*{Acknowledgements}

Thanks must go to the staff of the TBL and the AAT in their assistance in taking these data. This work has only been possible due to the brilliance of the late Meir Semel, from LESIA at Observatoire de Paris-Meudon. Sadly Meir passed away during 2012 but his memory and influence on spectropolarimetry will live on. The authors appreciate the time and dedication of the anonymous referee in producing constructive comments that has significantly improved this paper. This project has, in part, been supported by the Commonwealth of Australia under the International Science Linkages programme. S.V.J and S.BS acknowledge research funding by the Deutsche Forschungsgemeinschaft (DFG) under grant SFB 963/1 project A16. This project used the facilities of $\it{SIMBAD}$ and $\it{HIPPARCOS}$. This research has made use of NASA's Astrophysics Data System.

\label{lastpage}

\end{document}